\documentclass[a4paper,fleqn]{cas-sc}

\usepackage[authoryear]{natbib}

\graphicspath{{figs/}}  
\def\tsc#1{\csdef{#1}{\textsc{\lowercase{#1}}\xspace}}
\tsc{WGM}
\tsc{QE}
\tsc{EP}
\tsc{PMS}
\tsc{BEC}
\tsc{DE}

\usepackage{draftwatermark}
\SetWatermarkColor[gray]{0.9}
\SetWatermarkScale {0.7}
\SetWatermarkText {\textbf{Accepted version}}

\begin{document}


\let\WriteBookmarks\relax
\def\floatpagepagefraction{1}
\def\textpagefraction{.001}
\shorttitle{Numerical investigation of the wave force on a partially immersed rectangular structure}
\shortauthors{OI Gusev et~al.}

\title [mode = title]{Numerical investigation of the wave force on a partially immersed rectangular structure: Long waves over a flat bottom.}

\author{Oleg I. Gusev}[type=editor,
                        orcid=0000-0001-8110-5784]
\cormark[1]
\ead{gusev_oleg_igor@mail.ru}

\credit{Conceptualization, Software, Validation, Investigation, Writing - Original Draft, Writing - Review \& Editing, Visualization}

\address{Federal Research Center for Information and Computational Technologies, Academician M.A. Lavrentiev avenue, 6,
630090, Novosibirsk,
Russia}

\author{Gayaz S. Khakimzyanov}[
                        orcid=0000-0002-9837-7265]
												
\ead{gayaz.khakimzyanov@gmail.com}

\credit{Conceptualization, Methodology, Software, Validation,  Writing - Original Draft, Writing - Review \& Editing, Visualization}

\author{Leonid B. Chubarov}[
                        orcid=0000-0002-0169-1099]
												
\ead{chubarovster@gmail.com}

\credit{Conceptualization,  Writing - Original Draft, Writing - Review \& Editing, Project administration}

\cortext[cor1]{Corresponding author}

\begin{abstract}
This paper describes the results from a numerical estimation of the force exerted by long surface waves on a fixed and partially immersed rectangular structure.
The topic is connected with the need of making decisions on the design, placement and operation of hydrotechnical structures of this kind. The wave--structure interaction is simulated in the framework of a nonlinear potential flow model. The results obtained allow to determine the dependence of the hydrodynamic force on the length and amplitude of a single wave, the length and submergence of the structure.
We compare the results obtained for different initial wave shapes, and consider the differences between the forces acting on a vertical wall and a partially immersed structure. In particular, it is shown that at relatively small submergence of the structure, longer waves affect it less, while at large submergence and in the case of wave--wall interaction, the opposite behaviour is observed.
\end{abstract}

\begin{highlights}
\item The wave force on a structure is investigated using potential flow model
\item The force decreases at small submergence monotonically when wavelength increases
\item The dependence of the force on the wavelength is non-monotonic in other cases
\item Second local maximum appears in force chronograms at large wave amplitudes and structure submergence
\end{highlights}

\begin{keywords}
Long surface wave \sep Partially immersed structure \sep Wave force \sep Potential flow model \sep Numerical algorithm \sep Calculation results
\end{keywords}

\maketitle

\section{Introduction}

The design, placement and operation of high-tech objects anchored in coastal zones require an estimation of the force affected by long surface waves on these objects. The engineering calculation methods do not provide sufficient opportunities for such estimates in the case of complex hydrotechnical structures. Laboratory  modeling \citep[e.g.]{Kamynin_2010, Lu_Wang_2015} is complicated by the high cost and  laboriousness of experimental research, especially in the case studies with varying the parameters of the structure and wave. A reasonable alternative is to use numerical modeling methods, in particular, in the form of  hydrowave laboratory \citep{Nudner_2019}, which allows in some situations replace/expand laboratory modeling with numerical one, and thus reduce the cost and time of the project.

In this paper, the problem of determining the  force impact of long surface waves on a fixed and partially immersed rectangular structure is investigated.
Despite the simplicity of the geometry of a body and its immobility, the study of the interaction of waves with such objects is relevant due to the placement of high-tech objects of the similar shape in coastal zones (floating nuclear power plants, gas storage facilities, etc.).
Providing the safety of the population of coastal settlements and the environment certainly requires taking into account the possibility of a catastrophic force impact of the tsunamis \citep[e.g.]{Shokin_2019} on the hydrotechnical objects during the design and operation.

Even in this simple case, the problem has not yet been fully investigated, but some significant modeling results are known. For example, \cite{Mei_Black_1969} used the linear potential flow model to obtain analytical solutions for waves of infinitesimal amplitude and showed an increase in the reflected wave height with an increase in the length of a partially immersed body.
Theoretical studies of \cite{Bona_2005,Bresch_Lannes_2019,Lannes_2017} are devoted to the correctness of statements of the problem in the framework of the Boussinesq model.
In some articles, the problem of interaction of waves with fixed rectangular bodies was investigated numerically, but the computational results were presented very briefly, since only purpose was to demonstrate the capabilities of the proposed numerical algorithms.
For example, \cite{Lin_2006} provided information about the wave--body interaction for only one amplitude, and the results were presented as chronograms for only two virtual gauges.
\cite{Orzech_2016} considered cases for only two values of the length of a partially immersed body.

\cite{Engsig_2017} used numerical algorithms on unstructured grid to calculate the solitary wave runup on a fixed partially immersed body of rectangular cross-section. The results obtained in the framework of the model of potential fluid flows with a free boundary were given for only one amplitude of the incoming wave. \cite{Kamynin_2010} performed a more complete numerical study of the interaction of a solitary wave with a stationary rectangular body using a potential flow model.

The recent publication of \cite{Lu_Wang_2015} presents the results of a very detailed study of the interaction of a solitary wave with a fixed partially immersed body, performed using an ``integrated analytical--numerical'' approach. These authors investigated the dependence of amplitudes of reflected and transmitted waves on the parameters of the problem.
Using the finite-difference method, they solved one-dimensional (1D-) generalized Boussinesq equations in the outer region, in which the sought values are the free surface and the depth-averaged velocity potential. 
In the inner region beneath the structure, the 2D Laplace equation for the velocity potential with the boundary conditions of impermeability at the basin bottom and body bottom was solved analytically (using the spectral method).
At the bounds between the inner and outer regions, the values of potentials and their derivatives in the horizontal coordinate were equated.
The same problem, but without dividing into inner and outer regions, was investigated by \cite{Khakimzyanov_2002} and \cite{Chang_Wang_Hseih_2017} using the finite-difference method on curved meshes for the 2D potential flow model. 
\cite{Sun_etal_2015} performed simulations based on the finite element method for the same model to study the wave interaction with  a wall, a step on a bottom, a rectangular cylinder on free surface, etc. In particular, the dependence of the wave force on the cylinder submergence was examined for various wave amplitudes.  

In addition to numerical modeling, laboratory experiments \citep[e.g.]{Kamynin_2010,Lu_Wang_2015,Nudner_2017} were performed for various lengths and submergences of a partially immersed body and wave amplitudes. These studies have shown that the first two parameters strongly affect the wave patterns in front of the body and behind it, on the waves reflected from the body and passed behind it, on the wave loading on the front and back sides of the body.

Three-dimensional effects of the wave--body interaction were studied, for example, by \cite{Chang_2017} in the framework of a three-dimensional potential flow model. Unfortunately, these results were presented very briefly. \cite{Chen_Wang_2019} performed laboratory and numerical experiments investigating the wave interaction with a partially immersed and fixed vertical cylinder. The numerical results showed, in particular, that with larger submergence of the cylinder, the horizontal force increases, and the vertical force decreases. The same problem was investigated by \cite{Park_etal_2001} in the framework of potential flow and Navier--Stokes models. General good agreement of the computed results demonstrated that the viscous effects played a relatively
minor role in the problem under the study.

\cite{GK_DD_2020a} presented the statements of the problem within the hierarchy of mathematical models. Special attention was paid to the statements for shallow water models such as fully nonlinear weakly dispersive model and a non-dispersive shallow water model. The application of these models being on the lower levels of considered hierarchy assume splitting the flow region on the outer and internal parts. In the former part  there is a flow of water with a free boundary, while in the latter there is a flow in the space between the bottom of a partially immersed body and the bottom of the water area. To combine solutions of these parts, the compatibility conditions are used. 

In this paper, we focus on the force impact of long surface waves of various shapes on a partially immersed structures  in the framework of a 2D potential flow (Pot-) model. This model is widely used in studies of long surface waves, and its validation showed good results in many problems on the water flows \citep[e.g.]{Cooker_1997, Khakimzyanov_2001,Palagina_2019}. 
A finite-difference method with adaptive grids is implemented.  

In order to accurately estimate the force exerted by waves such as tsunamis, it is necessary to take into account many factors: uncertainties in the source parameters, wave transformation during propagation along an uneven bottom, diffraction, etc.
This will no longer be a methodological task, but a particular task, for a specific bay and source, without revealing any general patterns.
And the waveform that arrives at the bay can be very complex and not similar to the initial one. But analysing gauges, one can notice that the wave contains separate elements, pieces of gauge records, similar to a solitary wave, some long single waves, harmonic waves, N-waves, etc. 
To maintain a reasonable scope, this study will be limited to the effects of solitary and single waves propagating over a flat bottom.
We are clearly aware that the results of this work are rather fundamental, and we will present the results for more realistic tasks later and demonstrate the efficiency of the developed method.

The paper is structured as follows.
A brief description of the model equations and numerical algorithm is provided in Sections \ref{governing_equations} and \ref{algorithm}, respectively. 
Comparisons with the results of numerical calculations by other authors and experimental data are discussed in \ref{comparisons}.
Subsection \ref{general_view} gives general characteristics of the long wave interaction with a partially immersed structure. We further examine in \ref{main_results} the influence of the length and submergence of the structure, as well as the amplitude and length of the incident wave, on the value of the force on the structure. 

\section{Potential flow model}\label{governing_equations}

\begin{figure}[h!]
\centering
\parbox[t]{0.6\textwidth}{\centering {\it a}} \hfill \parbox[t]{0.35\textwidth}{\centering {\it b}}\\
\includegraphics[width=0.6\textwidth]{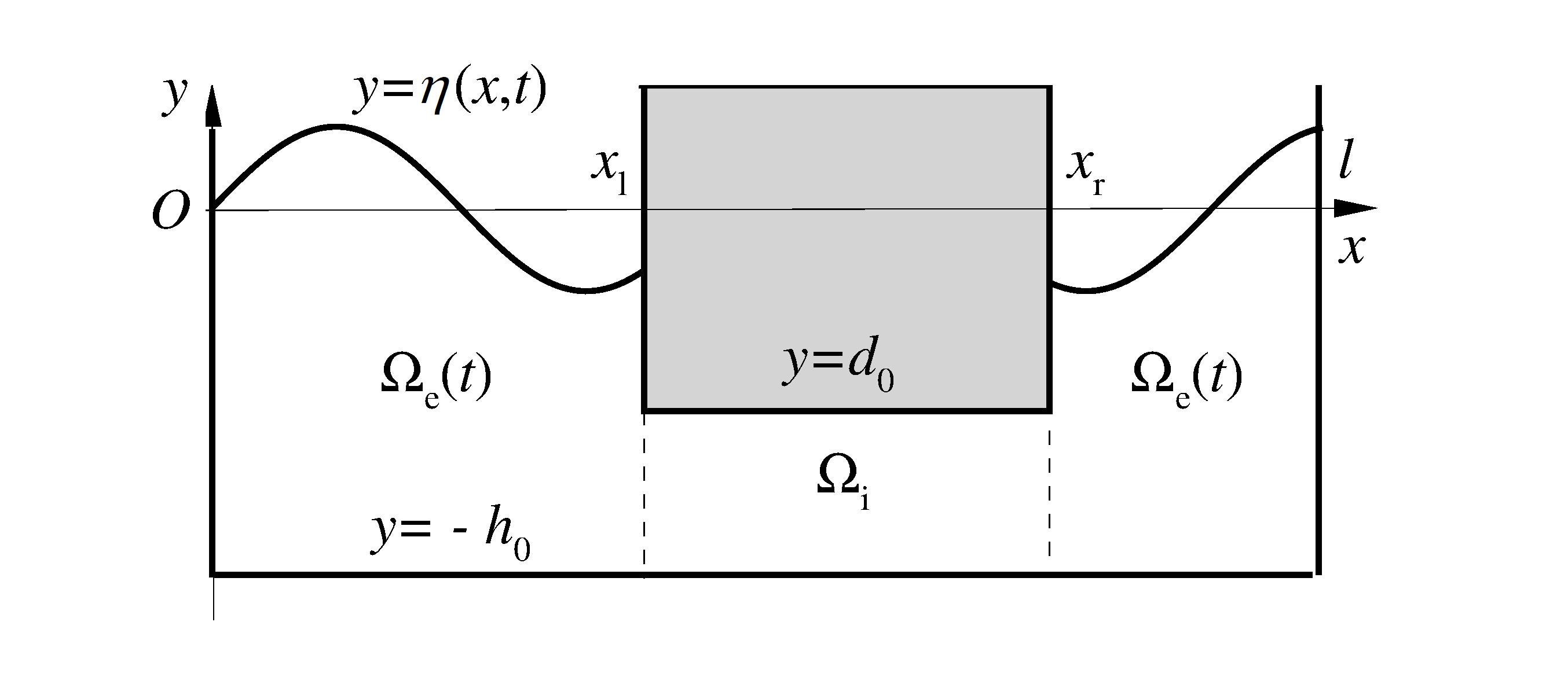}\hfill
\includegraphics[width=0.35\textwidth]{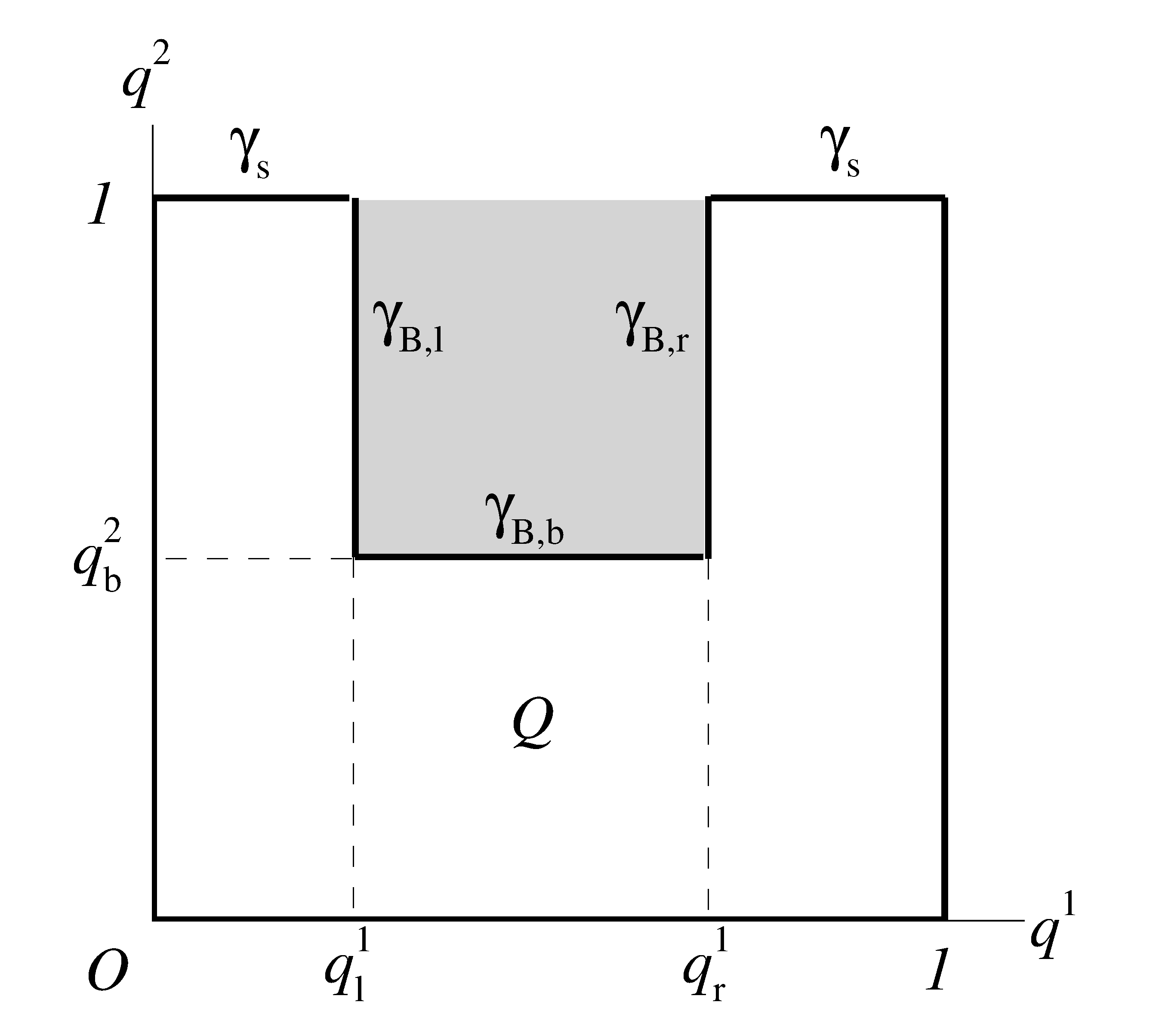}

{\caption{The sketch of the flow domain ({\it a}) and computational domain ({\it b}) in the problem of interaction of surface waves with a partially immersed fixed structure located in a basin with a horizontal bottom.}
\label{scheme_of_task}}
\end{figure}

In contrast to the study of \cite{GK_DD_2020a} which considered a three-dimensional statement of the problem in the Cartesian coordinate system $Ox_1x_2y$, we assume here that the flow parameters and geometry of the flow domain do not depend on one of the horizontal coordinates, for certainty from $x_2$.
Hereinafter, the notation $x$ is used for the first horizontal coordinate, $U(x, y,t)$ corresponds to the first velocity component of the Euler equation model, and the other notations do not differ from those used in \cite{GK_DD_2020a}.
It is assumed also that both the bottom of the basin and the bottom of the structure are horizontal, stationary, and are given by the equations $y=-h_0={\textrm {const}}$ and $y=d_0={\textrm {const}}$, respectively, $-h_0<d_0<0$. 
Thus, we consider a fixed partially immersed rectangular structure with vertical side faces located at the distances $x_{l}$ and $x_{r}$ from the left bound of the basin, $0<x_l<x_r<l$, where $x=0$ and $x=l$ are the coordinates of the left and right vertical walls of the basin.
The height of the side faces of the structure is assumed to be sufficient to exclude the  water flowing above them.
Under the assumptions made, the sketch of the flow domain looks as shown in figure~\ref{scheme_of_task}, where
\begin{equation*}
{\Omega}_e(t)=\left\{(x,y)\in \mathbb{R}^2\Big{|}\ x\in {\cal D}_e, \ -h_0\le y\le \eta(x,t)\right\},
\end{equation*}
\begin{equation*}
{\Omega}_i=\left\{(x,y)\in \mathbb{R}^2\Big{|}\ x\in {\cal D}_i, \ -h_0\le y\le d_0\right\},
\end{equation*}
${\cal D}_e=[0, x_l]\cup [x_r, l]$, $\ {\cal D}_i=(x_l, x_r)$,
$\ y=\eta(x,t)$ ($x\in {\cal D}_e$) is the free surface equation.

\subsection{Potential flow equations}

The equations of the 2D nonlinear Pot-model have the following form:
\begin{equation}
\Phi_{xx}+\Phi_{yy}=0\;,\qquad (x,y)\in \Omega(t)=\Omega_e(t)\cup\Omega_i\;,
\label{2D_11_1.14}
\end{equation}
\begin{equation}
\left(\eta_t+U \eta_x-V\right)\Big|^{y=\eta(x,t)} =0\;,\qquad {x}\in  {\cal D}_e\;,
\label{2D_11_1.5phi}
\end{equation}
\begin{equation}
\big(\Phi_t +\frac{U^2 + V^2}{2} +  g\eta\big)\Big|^{y=\eta( x,t)}=0\;,\qquad {x}\in  {\cal D}_e\;,
\label{2D_11_1.16}
\end{equation}
where $g$ is the acceleration of gravity,
\begin{equation}
U=\Phi_x\;, \quad V=\Phi_y\;,\qquad (x,y)\in \Omega(t)\;.
\label{2D_11_1.1.71}
\end{equation}
These equations are supplemented by the impermeability conditions at the bottom
\begin{equation}
\Phi_y\Big|_{y=-h_0}=0\;,\qquad {x}\in {\cal D}_e\cup{\cal D}_i
\label{2D_11_1.15}
\end{equation}
and the sides of the structure
\begin{equation}
\Phi_y\Big|_{y=d_0} =0\;,\qquad {x}\in {\cal D}_i\;,
\label{2D_11_bodybottom_2}
\end{equation}
\begin{equation}
\Phi_x=0\,, \qquad x\in \left\{x_l, x_r\right\}\;,\quad  d_0\le y\le \eta(x,t)\;.
\label{2D_Gamma}
\end{equation}
The free passing of surface waves outside the considered area is modeled by non-reflecting boundary conditions at the boundaries of the basin \citep{Palagina_2019}.

To calculate the pressure $p$ in potential flows of fluid with density $\rho={\textrm {const}}$, the Cauchy--Lagrange integral is used
\begin{equation}
\frac{p(x,y,t)}{\rho}=-\Big(\Phi_t(x,y,t) +\frac{1}{2}U^2(x,y,t) +\frac{1}{2}V^2(x,y,t)+   gy\Big)\;,\qquad  (x,y)\in \Omega(t)\;,
\label{2D_press_P}
\end{equation}
to determine the pressure at any point, in particular, at  the front, back and bottom faces of the structure. Using these pressure values, the force acting on the faces can be found, as well as the total force ${\vec {\cal F}}$ acting on the entire body:
\begin{equation}
{\vec {\cal F}}(t)=-\iint\limits_{S(t)}p(x,y,t){\vec n}(x,y)dS,
\label{Force_pot_fluids}
\end{equation}
where ${\vec n}$ is an exterior normal to the side faces of the body and its bottom, $S(t)$ is the ``wetted'' surface of the structure. The horizontal component $F$ of this force ${\vec {\cal F}}$ is expressed by the formula
\begin{equation}
F(t)=\int\limits_{d_0}^{\eta(x_l,t)}p(x_l,y,t)dy-\int\limits_{d_0}^{\eta(x_r,t)}p(x_r,y,t)dy.
\label{2D_Force_1}
\end{equation}

\subsection{Initial conditions}

It is necessary for the Pot-model  to set the initial conditions for the free surface and the velocity vector field:
\begin{equation}
\eta(x, 0)=\eta_0(x), \quad U(x,y,0)=U_0(x,y), \quad V(x,y,0)=V_0(x,y).
\label{initial_PFK}
\end{equation}

As noted in the Introduction, two types of tsunami wave components will be analysed here, corresponding to a solitary wave (of infinite length) and a single wave of finite length. In both cases, the initial field of the velocity vector is a potential one, so the initial values for the potential $\Phi (x,y,0)$ can be  determined uniquely \citep{GK_DD_2018e}.

Unfortunately, the exact solution in the form of solitary wave for the Pot-model is unknown, thus we use the initial conditions in the form of solitary wave for the Serre--Green--Naghdi equations, which are given by the following formulas \citep{Palagina_2019}:
\begin{equation}
  \eta_0(x)=a_0{\textrm {sech}}^2(X)\;,
  \label{Full_an_sol_eta}
\end{equation}
\begin{equation}
\begin{array}{c}
\displaystyle
U_{0}(x,y)  =  u_0(x) 
\left[1+\Big(\frac{1}{4}-\frac{3}{4}\frac{(y+h_0)^2}{H_0^2(x)}\Big)
\frac{H_0(x)\big(2a_0-3\eta_0(x)\big)+ 4\big(\eta_0(x)-a_0\big)\eta_0(x)}{h_0(a_0+h_0)}\right]\;,\\[5mm]
\displaystyle
V_{0}(x,y)  =  \sqrt{3a_0g}\;\frac{\eta_0(x)}{H_0^2(x)}\;(y+h_0) \tanh(X)\;, \quad -h_0\le y\le \eta_0(x)\;,
\end{array}
\label{init_cond_6_new}
\end{equation}
where $H_0(x)=h_0+\eta_0(x)$, $X= k(x-x_0)$, $a_0$~is the amplitude of the initial wave,  $x=x_0$~is the position of its crest,
\begin{equation}
c_0=\sqrt{g(a_0+h_0)}, \quad  k=\frac{1}{h_0}\sqrt{\frac{3a_0}{4(a_0+h_0)}}, \quad u_0(x)=c_0\frac{\eta_0(x)}{H_0(x)}\;.
\label{Full_an_sol_u}
\end{equation}
The numerical calculations of \cite{Khakimzyanov_2001}  showed that the initial data (\ref{Full_an_sol_eta}), (\ref{init_cond_6_new}) in the Pot-model  generate at $t>0$ a solitary wave moving to the right  at a constant speed, and for small enough  amplitudes, the shape of such a wave differs slightly from the specified with (\ref{Full_an_sol_eta}).

To determine the dependence of the characteristics of the wave--structure interaction on the shape and length of the wave, we will further consider the so-called single waves, which have a finite pre-set length and, as with solitary waves, a single crest.
The single wave of length $\lambda>0$ is defined by the following formulas \citep{Palagina_2019}:
\begin{equation}
\eta_0(x)=\left \{\begin{array}{cc}
\displaystyle
\frac{a_0}{2} \Big( 1+\cos (X) \Big),  & \displaystyle \left|x-x_0\right|\le {\lambda}/{2}\;,\\
0, & \displaystyle \left|x-x_0\right|>{\lambda}/{2}\;,
\end{array} \right.
\label{test_eta_0}
\end{equation}
\begin{equation}
\begin{array}{c}
\displaystyle
U_{0}(x,y)  =  u_0(x) + c_0\frac{h_0}{H_0(x)}\Big(\frac{1}{3}-\frac{(y+h_0)^2}{H_0^2(x)}\Big)
\frac{H_0(x)\big(a_0-2\eta_0(x)\big)+ 4\big(\eta_0(x)-a_0\big)\eta_0(x)}{\left(\lambda/\pi\right)^2}\;,\\[5mm]
\displaystyle
V_{0}(x,y)=c_0\frac{\sqrt{\big(a_0-\eta_0(x)\big)\eta_0(x)}}{H_0^2(x)}\;\frac{2\pi h_0}{\lambda}\;(y+h_0)\; {\textrm {sgn}}(X) \;, \quad -h_0\le y\le \eta_0(x)\;.
\end{array}
\label{init_cond_single_wave}
\end{equation}
The same notation as in formulas (\ref{Full_an_sol_eta})--(\ref{Full_an_sol_u}) are used here, except that $k=2\pi/\lambda$.

\begin{figure}[h!]
\centering
\includegraphics[width=0.95\textwidth]{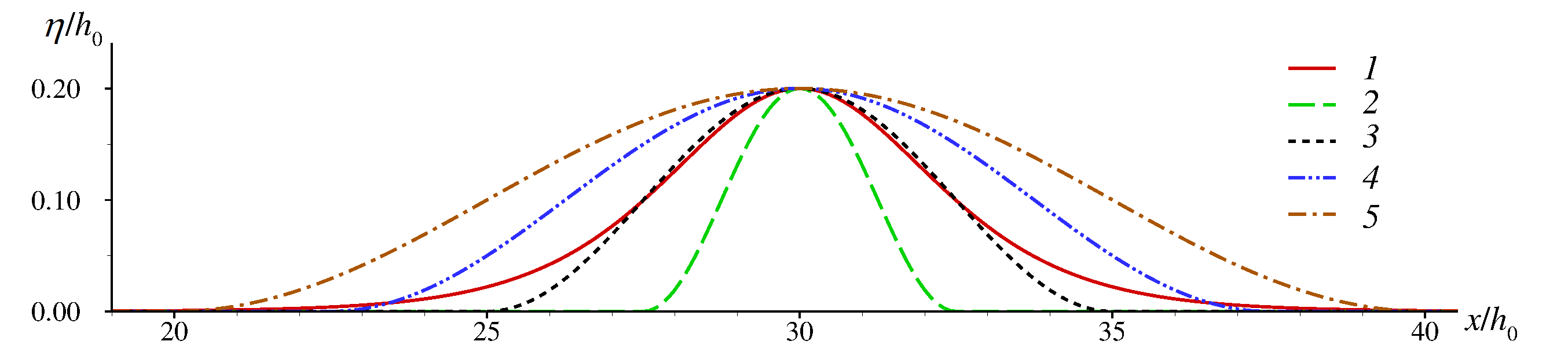}
{\caption{The initial free surface of the solitary ({\sl 1}) and single ({\sl 2})--({\sl 5}) waves of $a_0/h_0=0.2$ and $\lambda/h_0=5$~({\sl 2}), $10$~({\sl 3}), $15$~({\sl 4}), $20$~({\sl 5}).}
\label{Soliton+finite_wave}}
\end{figure}

The figure~\ref{Soliton+finite_wave} shows the profiles of the initial free surface for the solitary wave and the single waves of different lengths. It can be seen the shape of long single waves differs significantly from the shape of a solitary wave of the same amplitude.
\begin{figure}[b!]
\centering
\parbox[t]{0.45\textwidth}{\centering {\it a}} \hfill \parbox[t]{0.45\textwidth}{\centering {\it b}}\\
\includegraphics[width=0.45\textwidth]{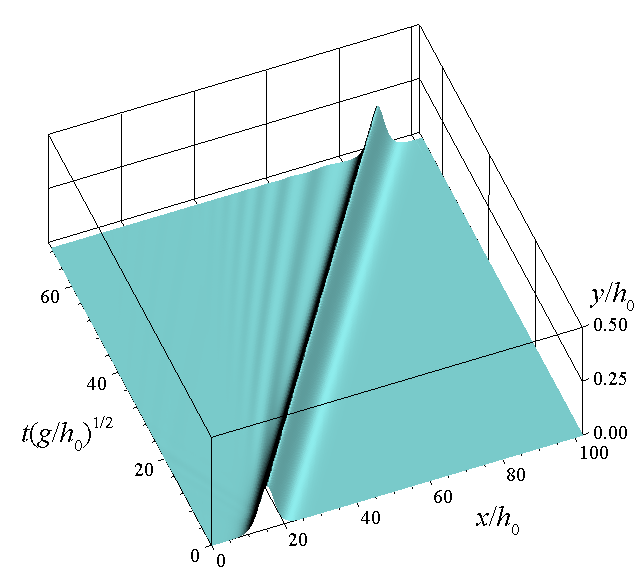}\hfill
\includegraphics[width=0.45\textwidth]{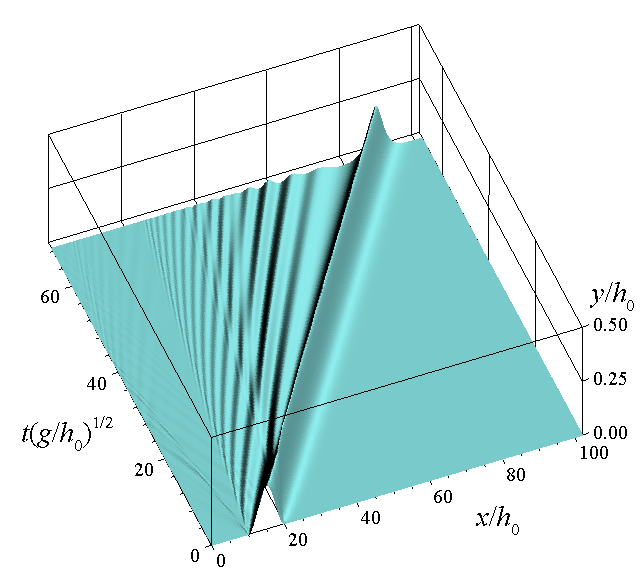}

{\caption{Propagation above the horizontal bottom of the solitary ({\it a}) and single waves of length $\lambda/h_0=10$ ({\it b}).} $a_0/h_0=0.2$, $x_0/h_0=15$. 
\label{Surface_with_two_waves}}
\end{figure}

Figure~\ref{Surface_with_two_waves} shows the process of wave propagation above the horizontal bottom. The initial solitary wave transforms at $t>0$ into soliton-like wave with a small dispersion tail  (Fig.~\ref{Surface_with_two_waves},~{\it a}).
Such dispersion tail for the single wave is more pronounced (Fig.~\ref{Surface_with_two_waves},~{\it b}). Its presence in the wave pattern demonstrates a fairly long process of forming a soliton from the initial wave which does not correspond to it \citep[e.g.]{Pelinovsky_1996}.
This effect observed in the computations leads to the conclusion that for numerical investigation of the force affected by the single waves on a partially immersed structure, the corresponding initial free surface should be placed in close proximity to the frontal face of the structure, while satisfying the condition
\begin{equation}
\frac{\lambda}{2}\;\le \;x_0 \;\le\;  x_l- \frac{\lambda}{2}\;.
\label{x_0_under_t0}
\end{equation}
The same condition should be satisfied for the initial solitary wave. To check this, we define the ``effective'' length of the initial solitary wave as the distance between two points of free surface that are symmetrical with respect to the crest, in which the value of the wave height is equal to 10\% of the amplitude $a_0$. This is not the only way to determine the ``effective'' wave length. At the initial moment of time, a state of rest is set under the structure and to its right.

\section{Numerical algorithm}\label{algorithm}
For numerical investigation of the problem of the wave--structure interaction, we adopt the modification of the algorithm proposed in studies of \cite{Khakimzyanov_2001,GK_DD_2018e,Palagina_2019} to calculate potential flows of fluid with a free surface in basins with movable or stationary walls and bottom fragments. This algorithm is based on the use of  the problem statement transformed for  the following moving curvilinear coordinate system:
\begin{equation}
x=x(q^1, q^2, t), \quad y=y(q^1, q^2, t),
\label{2.1}
\end{equation}
in which all the bounds of the domain ${\Omega}(t)$ lie on the coordinate lines of the first or second family. The time dependent coordinate transformation (\ref{2.1}) is assumed to be a diffeomorphism from the reference domain ${\Omega} (t)$ to the~fixed computational domain $Q$ lies at  the plane of variables $q^1$, $q^2$. In contrast to the study of \cite{GK_DD_2018e} and  \cite{Palagina_2019}, the computational domain used here is a unit square with a rectangle cut out from the top.

Let's assume that the side bounds of the computational domain $Q$ are mapped by the transformation (\ref{2.1}) to the vertical walls of the basins  (Fig.~\ref{scheme_of_task}), the lower bound is mapped to the bottom of the basin, the sides of the excised rectangle are mapped to the vertical sides and the bottom of the structure, respectively,
\begin{equation*}
\begin{array}{c}
\displaystyle
\gamma_{B,l}=\left\{ {\vec q}\big|\; q^1=q^1_l,\, q^2_b\le q^2\le 1\right\},\quad
\gamma_{B,r}=\left\{{\vec q}\big|\; q^1=q^1_r, \, q^2_b\le q^2\le 1\right\},\\[2mm]
\displaystyle
\gamma_{B,b}=\left\{{\vec q}\big|\; q^1_l\le q^1\le q^1_r,\, q^2=q^2_b\right\}
\end{array}
\end{equation*}
Here
${\vec q}=(q^{1}, q^{2})$, $\ 0<q^1_l<q^1_r<1$, $\ 0<q^2_b<1$.
In the new coordinates, the free surface
\begin{equation*}
\gamma_s=\left\{ {\vec q}\big|\; 0\le q^1\le q^1_l,\, q^2=1\right\}\cup \left\{ {\vec q}\big|\; q^1_r\le q^1\le 1,\, q^2=1\right\}
\end{equation*}
is fixed and consists of the combination of two line segments lying on the upper side of the unit square. The assumption $q^2_b<1$  is essential  for the algorithm, which means that the bottom of the partially immersed structure is under water at every time moment.

The unit square in the $ Oq^1q^2 $ plane  (see Fig.~\ref{scheme_of_task}, ({\it b})) is covered by a rectangular grid  with nodes $ {\vec q}_{\vec j} $ and steps $h_{1} = 1 / N_1 $ and $ h_{2} = 1 / N_2 $ in the direction of the $Oq^{1} $ and $Oq^2$, respectively, where ${\vec j} = (j_1, j_2)$, $ j_1 = 0, \ldots, N_1 $, $j_2 = 0, \ldots, N_2$.
The boundaries $ \gamma_{B, l} $, $ \gamma_{B, r} $, $ \gamma_{B, b} $  lie on the coordinate lines of the grid, i.e. $ q^1_l = j_lh_1 $, $ q^1_r = j_rh_1 $, $ q^2_b = j_bh_2 $, where $ 0<j_l <j_r <N_1 $, $ 0 <j_b <N_2 $.
Laplace equation (\ref {2D_11_1.14}), kinematic (\ref {2D_11_1.5phi}), dynamic (\ref {2D_11_1.16}) and boundary  conditions (\ref {2D_11_1.15})--(\ref {2D_Gamma}) are written in the new coordinate system and are approximated only at those grid nodes that belong to the computational domain $Q$.
So, in the new coordinates, the Laplace equation (\ref{2D_11_1.14}) takes the form of a second-order elliptic equation with variable coefficients
\begin{equation}
\frac{\partial}{\partial q^{1}}\left ( k_{11} \frac{\partial \Phi}{\partial q^{1}}+ k_{12} \frac{\partial \Phi}{\partial q^{2}} \right )+
\frac{\partial}{\partial q^{2}}\left ( k_{21} \frac{\partial \Phi}{\partial q^{1}}+ k_{22} \frac{\partial \Phi}{\partial q^{2}} \right ) = 0\;,
\quad {\vec q}\in Q\;,
\label{2.34}
\end{equation}
and boundary conditions  (\ref{2D_11_1.15})--(\ref{2D_Gamma}) can be written as
\begin{equation}
\left(k_{21}\frac{\partial\Phi}{\partial q^1}+k_{22}\frac{\partial\Phi}{\partial q^2}\right) \biggm|_{{\vec q}=(q^1, 0), \ q^1\in[0, 1]}=0\;,
\label{Pal_contr_5a}
\end{equation}
\begin{equation}
\left(k_{21}\frac{\partial\Phi}{\partial q^1}+k_{22}\frac{\partial\Phi}{\partial q^2}\right) \biggm|_{{\vec q}\in \gamma_{B,b}}=0\;,
\label{Pal_contr_5}
\end{equation}
\begin{equation}
\left(k_{11}\frac{\partial\Phi}{\partial q^1}+k_{12}\frac{\partial\Phi}{\partial q^2}\right) \biggm|_{{\vec q}\in \gamma_{B,l}\cup \gamma_{B,r}}=0\;,
\label{Pal_contr_6}
\end{equation}
where
\begin{equation}
k_{11}=\frac{g_{22}}{J}, \quad k_{12}=k_{21}=-\frac{g_{12}}{J}, \quad k_{22}=\frac{g_{11}}{J},
\label{koef_Phi}
\end{equation}
\begin{equation}
g_{11}= x^2_{q^1}+y^2_{q^1}, \quad g_{12}=g_{21}=x_{q^1}x_{q^2}+y_{q^1}y_{q^2}, \quad g_{22}= x^2_{q^2}+y^2_{q^2},
\label{geom_characters}
\end{equation}
$J=x_{q^1}y_{q^2}-x_{q^2}y_{q^1}$ is the Jacobian of transformation (\ref{2.1}),  $J>0$.

The finite difference equations for the velocity potential $\Phi$ can be obtained using the integro-interpolation method \citep{GK_DD_2018e} with the 9-points template for internal grid nodes. 
At the boundary nodes, the boundary conditions are approximated using (\ref{2.34}).
Moreover, the boundary conditions (\ref{Pal_contr_5a})--(\ref{Pal_contr_6}) are used  in the integro-interpolation method in exact differential form. 
As a result, the finite-difference equations of the second approximation order on six or eight-point templates are constructed at the boundary nodes, and the resulting difference problem has a number of positive properties \citep{GK_DD_2018e}.

The algorithm proposed here has peculiarities generated by the introduction into the computational domain of a partially immersed structure. Let's consider  some of them.
Suppose that for $t=t^n$ the curvilinear mesh ${\vec x}^n$ was constructed and the values $\eta^n$ and $\Phi^n$ were computed. 
Calculation of the solution $\eta^{n + 1}$, $\Phi^{n+1}$ on the $(n+1)$-th layer in time, i.e. at time $t^{n+1}=t^{n}+\tau_n$, where $\tau_{n}$~is a time step, begins with determining the values of the potential $\Phi_{\vec j}^{n + 1}$ at grid nodes lying on the pre-image of the free boundary.
We use the difference approximation of condition (\ref{2D_11_1.16}) written in the new coordinates for this purpose.  

In contrast to the algorithm presented in \cite{GK_DD_2018e, Palagina_2019}, the pre-image $\gamma_s$ of the free boundary here consists of two segments, and non-reflecting conditions are used on the side walls of the basin. Therefore, the values of $\Phi_{\vec j}^{n+1}$ need to be calculated only in the nodes ${\vec q}_{\vec j}$ with numbers $j_1\in \{1,\ldots, j_l\}\cup \{j_r,\ldots,N_1-1\}$, $j_2=N_2$.
Then the non-reflective conditions \citep{Palagina_2019} are applied and the new values $\Phi_{\vec j}^{n+1}$ are calculated at all nodes of the left and right boundaries  of the computational domain $Q$.
The computed values $\Phi_{\vec j}^{n+1}$ are used as discrete Dirichlet boundary conditions for solving the above-mentioned difference problem of determination $\Phi_{\vec j}^{n+1}$ at the internal nodes of domain $Q$ and at nodes belonging to the pre-images of the bottom of the basin and to the faces of the structure (except for the nodes ${\vec q}_{0, 0}$, ${\vec q}_{N_1,0} $, $ {\vec q}_{j_l, N_2}$ and ${\vec q}_{j_r,N_2} $, where the new  values of the potential have already been calculated).
Similar to the study of \cite{GK_DD_2018e}, we adopt an iterative Successive Over–Relaxation method to solve the difference problem.

Next, the new position of the free surface $\eta_{j_1}^{n+1}$ ($j_1\in \{1,\ldots, j_l\}\cup \{j_r,\ldots,N_1-1\}$) is determined using approximation of kinematic condition (\ref{2D_11_1.5phi}) written in the new coordinates \citep{Khakimzyanov_2001}. The values $\eta_{0}^{n+1}$ and $\eta_{N_1}^{n+1}$ are calculated at the boundary nodes using non-reflecting conditions \citep{Palagina_2019} .

All the algorithm fragments presented above are executed on the grid ${\vec x}_{\vec j}^n$ corresponding to the $n$-th time layer ($t=t^n$). 
The algorithm is completed by calculating the grid ${\vec x}_{\vec j}^{n+1}=\left(x_{\vec j}^{n+1}, y_{\vec j}^{n+1}\right)$ for the next time layer.
Compared to the study of \cite{GK_DD_2018e}, we use simpler computational grids with fixed vertical coordinate lines of the second family, i.e. with time-invariant node abscissae.
Moreover, the grid does not change with time for $ y\le d_0 $, it is rectangular, uniform along the $Oy$ axis.
The grid is uniform also in the horizontal direction in the region $\Omega_i $ under the structure (see Fig.~\ref{scheme_of_task}): $\Delta x=L/(j_r-j_l)$.
The grid is movable only in the outer region $\Omega_e$ for $y> d_0$, and its nodes move only in the vertical direction, adapting to the movable free surface.
The mesh can be refined in this region in the horizontal direction, in order to improve the accuracy of calculations in the vicinity of the structure.

When the new mesh is constructed, we repeat the calculations to match the values of $\Phi^{n+1}$ and $\eta^{n+1}$ with mesh ${\vec x}_{\vec j}^{n+1} $, its geometric characteristics (\ref{geom_characters}) and coefficients (\ref {koef_Phi}). Some details of this recalculation are described in \cite{GK_DD_2018e}.

The wave force  impact on a partially immersed structure is also calculated at each time step.
Formula (\ref{2D_press_P}) for determining the pressure takes the following form in the new coordinate system:
\begin{equation}
\frac{p({\vec q},t)}{\rho}=-\Big(\Phi_t({\vec q},t) -\left(Ux_t+Vy_t\right)({\vec q},t)+\frac{1}{2}\left(U^2 +V^2\right)({\vec q},t)+   gy({\vec q},t)\Big)\;,\qquad  {\vec q}\in Q\;,
\label{press_p_q1q2}
\end{equation}
where the Cartesian velocity components are calculated through the potential $\Phi$ by the formulas
\begin{equation*}
U({\vec q},t)=\frac{\Phi_{q^1}y_{q^2}-\Phi_{q^2}y_{q^1}}{J}({\vec q},t), \quad
V({\vec q},t)=\frac{-\Phi_{q^1}x_{q^2}+\Phi_{q^2}x_{q^1}}{J}({\vec q},t), \qquad {\vec q}\in Q\; .
\end{equation*}
An analogue of formula (\ref{2D_Force_1}) for calculating the force horizontal component has the following form:
\begin{equation}
F(t)=\int\limits_{q^2_b}^{1}p(q^1_l,q^2,t)dq^2-\int\limits_{q^2_b}^{1}p(q^1_r,q^2,t)dq^2\;.
\label{2D_Force_2_q1q2}
\end{equation}
We adopt the rectangular quadrature rule to calculate the integrals in expression (\ref{2D_Force_2_q1q2}).

\section{Results and discussion}

This section provides the estimates of the dependence of  computed wave characteristics on the parameters of the wave and structure, as well as the results of comparing the obtained characteristics with the available experimental data and with the results determined by other authors.

\subsection{Comparisons with experimental data and results of other authors}\label{comparisons}

\begin{figure}[h!]
\centering
\includegraphics[width=0.49\textwidth]{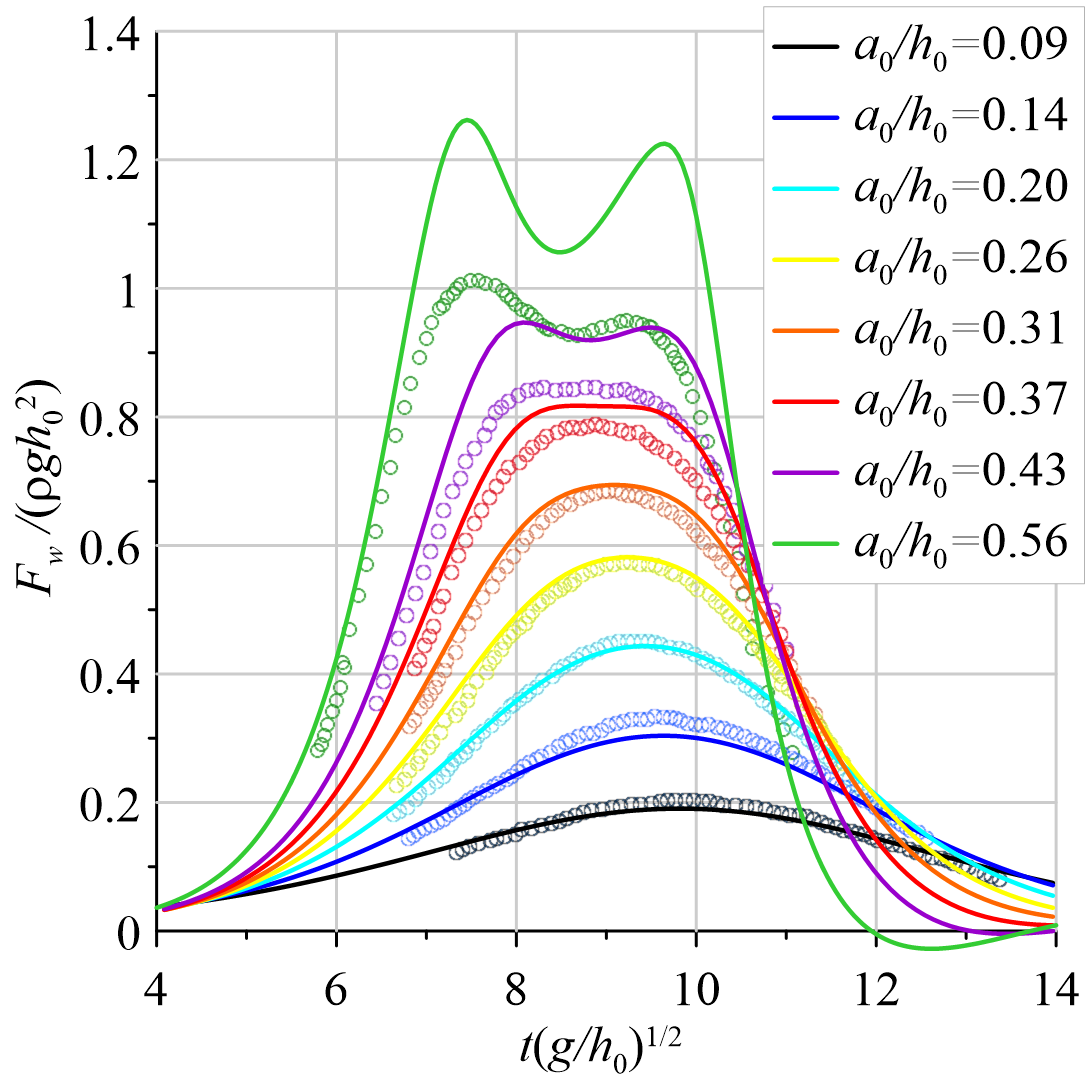}

{\caption{Chronograms of the force acting  on the fixed vertical wall for the waves of different amplitudes: computed in this study (solid lines) and defined in the laboratory  experiments of  \cite{Chen_et_al_2019} (circles).
}
\label{Chen_exp_chr}}
\end{figure}

\begin{figure}[h!]
\centering
\includegraphics[width=0.49\textwidth]{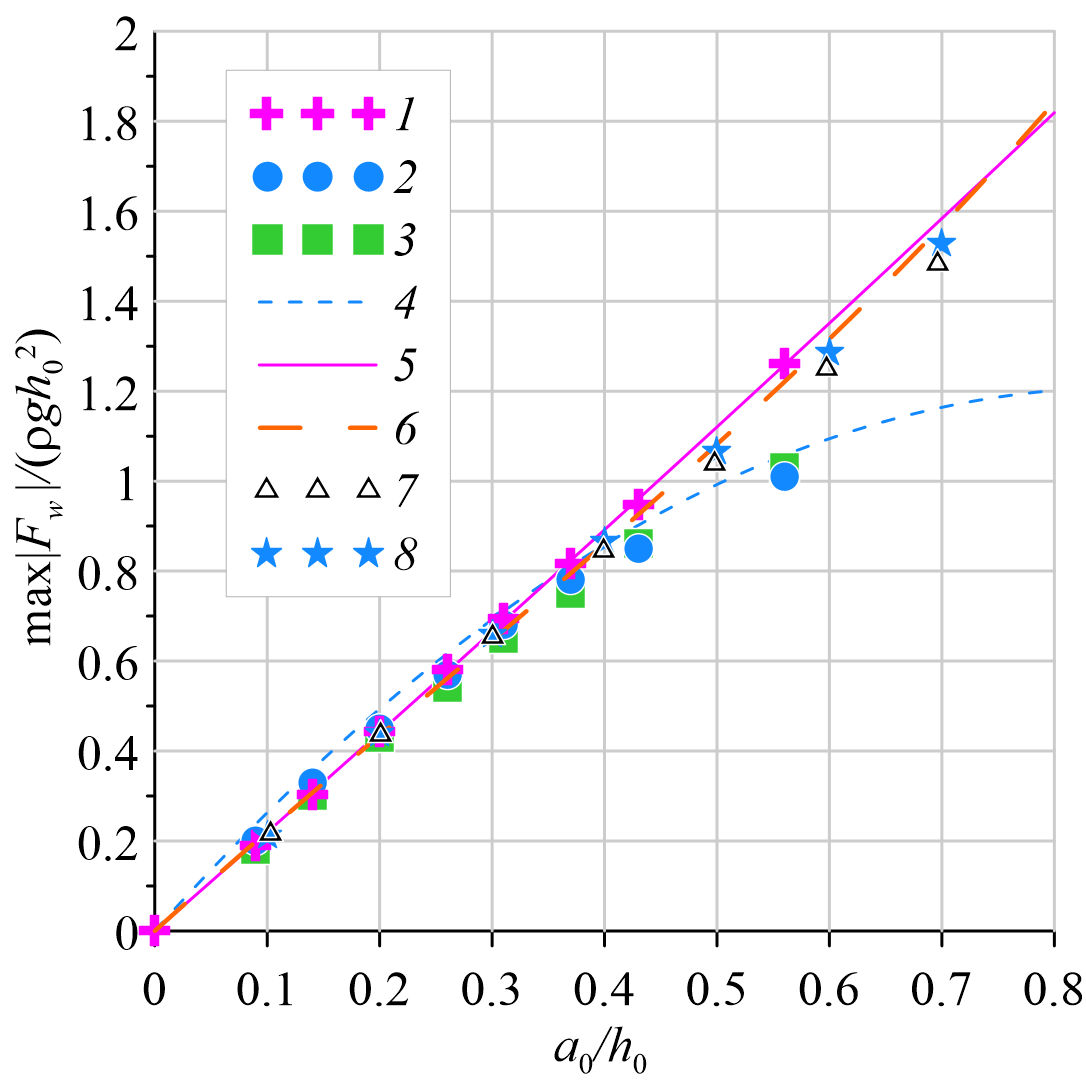}
{\caption{Dependences of the maximum (over the entire time of calculations) wave force acting on the fixed vertical wall on the amplitude of the incident wave: {\it 1}, Pot-model solutions; {\it 2}, experimental data \cite {Chen_et_al_2019}; {\it 3}, Navier--Stokes model solutions of \cite{Chen_et_al_2019}; {\it 4}, curve corresponding to the formula \eqref{Chen} approximating the data {\it 2} (after subtracting the hydrostatic pressure component); {\it 5},  curve corresponding to the formula \eqref{Fw_a_PF}, which approximates the computed data {\it 1}; {\it 6} and {\it 7}, results of calculations from the studies of \cite {Cooker_1997} and \cite {Zheleznyak_1985}, respectively; {\it 8}, solutions obtained by the nonlinear dispersion model \citep{GK_DD_2020a}.
}
\label{Chen_exp_max}}
\end{figure}

In this section we present the results of comparing the wave characteristics computed using the above mathematical model and numerical algorithm, with the results of laboratory and computational experiments performed by other authors. Unfortunately, there is an obvious lack of publications related to experimental studies of the wave force acting on fixed partially immersed bodies. In contrast, the results of numerical and laboratory studies of the wave force acting  on a vertical wall are well known, i.e. the recently published study of \cite{Chen_et_al_2019}, where the materials of small-scale (with a layer of unperturbed liquid 15~cm thick) laboratory experiments for the waves of different amplitudes are presented. The paper  \cite{Chen_et_al_2019} contains chronograms of the force, as well as the dependence of the maximum force (for the entire time of the experiment) on the wave amplitude, presented by these authors also in the form of analytical formula approximating the experimental data,
\begin{equation}
\frac{\max{|F_w|}}{\rho g h_0^2}=-1.61\left(\frac{a_0}{h_0}\right)^2+2.79\left(\frac{a_0}{h_0}\right)+0.5,
\label{Chen}
\end{equation}
where $F_w(t)$ is the force acting on the vertical wall. 
The hydrostatic component in a dimensionless form equal to $ 0.5 $ is subtracted from the corresponding data by \cite{Chen_et_al_2019} for correctness of the comparisons (see Fig.~\ref{Chen_exp_chr}, \ref{Chen_exp_max}). The chronograms in Fig. \ref{Chen_exp_chr} are shifted along the time axes in order their initial positions coincide. It is necessary since in laboratory experiments the wave generator was located at a distance of about $ 60h_0 $ from the vertical wall, and in the numerical experiments of this study, the initial position of the solitary wave front was $ 10h_0 $ from the wall.

Figures~\ref{Chen_exp_chr} and \ref{Chen_exp_max} demonstrate a good agreement between the computed  results and experimental data for $a_0/h_0 \le 0.31 $. The excess of the experimental data by the force characteristics calculated using the Pot-model in the region of large amplitudes may be connected with the viscosity, which turned out to be a significant factor in the small-scale laboratory experiments and was not taken into account by the Pot-model. These statements  are confirmed by the closeness of the results of calculations performed by \cite {Chen_et_al_2019} using the Navier--Stokes model and the corresponding experimental data.
Analysis of the trends shown in Figure~\ref{Chen_exp_max} by the curve {\sl 4}, which approximates the experimental data of \cite{Chen_et_al_2019}, and the curve {\sl 5}, which approximates by the formula
\begin{equation}
\frac{\max{|F_w|}}{\rho g h_0^2}=0.11\left(\frac{a_0}{h_0}\right)^2+2.19\left(\frac{a_0}{h_0}\right)
\label{Fw_a_PF}
\end{equation}
the numerical results obtained using the Pot-model show that the dependence \eqref {Chen} slightly overestimates the experimental data, and the curve {\sl 5} is quite close, including in the region of large amplitudes, to the results of calculations published in the papers of  \cite{Cooker_1997} and \cite {Zheleznyak_1985} obtained using the potential flow model and the nonlinear dispersion model, respectively. Also, the solutions obtained using the Pot-model  agree well with the results of calculations performed with the fully nonlinear dispersion model \citep{GK_DD_2020a}.

\begin{figure}[h!]
\centering
\includegraphics[width=0.49\textwidth]{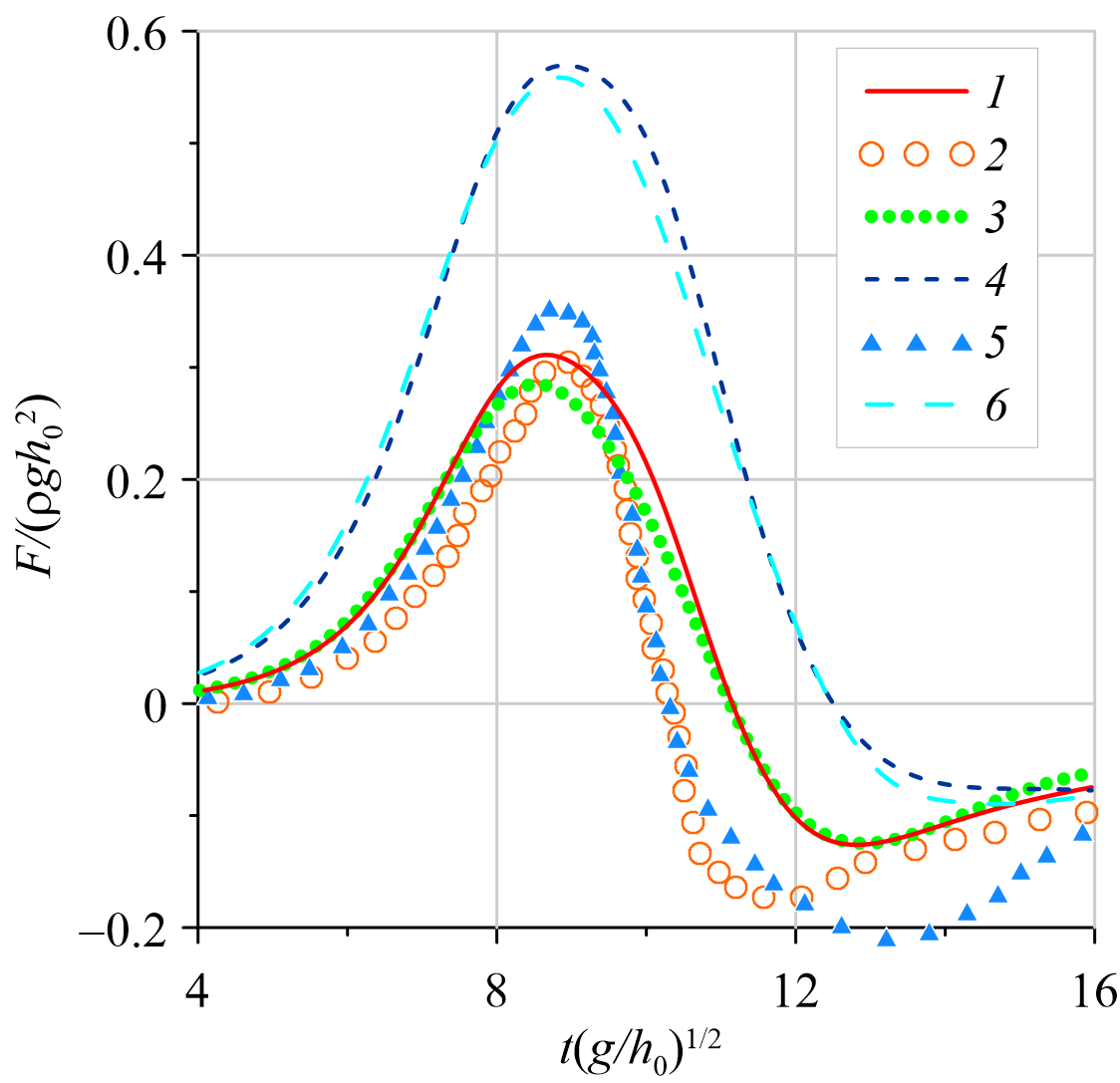}
\\
{\caption{Chronograms of the total horizontal component of the wave force acting on the body immersed to the depth of $ d_0 / h_0 = -0.5 $ ({\it 1, 2, 3}) and $ d_0 / h_0 = -0.9 $ ({\it 4 , 5, 6}): {\it 1, 4}, Pot-model solutions of the present study; {\it 2, 5}, calculation results of  \cite{Lu_Wang_2015}; {\it 3, 6},  solutions obtained by the nonlinear dispersion model \citep{GK_DD_2020a}.
$a_0/h_0=0.3$, $L/h_0=4$.
\label{Lu}}}
\end{figure}

The materials published in the paper of \cite{Lu_Wang_2015} allow to compare the solutions obtained with the Pot-model with the results of numerical and laboratory experiments on the waves interaction with a fixed and partially immersed body. The data of small-scale laboratory experiments are presented only by chronograms, determined on both sides of the body at certain distances from it. Unfortunately, the data of the force measurements have not been published by these authors. Therefore, we compare the results of calculations of the horizontal component of the total wave force  obtained here using the Pot-model with the results by the ``integrated numerical--analytical'' model \cite{Lu_Wang_2015} (see Fig.~\ref{Lu}) only. The  compared results were computed for two values of the body submergence: $ d_0 / h_0 = -0.5 $ and $ d_0 / h_0 = -0.9 $, amplitude $ a_0 / h_0 = 0.3 $ and body length $ L / h_0 = 4 $.
The digitized values of the study \cite{Lu_Wang_2015} are transformed here in a dimensionless form, taking into account the used basin depth $h_0 = 0.0762$~m.

Comparisons of the calculated and experimental chronograms of the free surface measured at a distance from the body showed their good agreement (these results are not presented). In this study, we present the comparisons of the force action characteristic computed using these two different mathematical models and different numerical algorithms on Fig.~\ref{Lu} which shows an excellent agreement of the results for the body submergence $ d_0 / h_0 = -0.5 $ and a significant difference when the thickness of the fluid layer under the body decreases ($ d_0 / h_0 = -0.9 $). This result indicates the limited range of applicability of the ``integrated numerical--analytical'' model of \cite{Lu_Wang_2015} derived from the more complete Pot-model using additional simplifying assumptions on the flow characteristics. It should be noted also that the revealed limitation of the Pot-model affects at a relatively small submergences of the body and large wave amplitudes (e.g. $ d_0 / h_0 = -0.1 $ and $ a_0 / h_0 = 0.3$), when the bottom of the body is drained. For this reason, the comparisons results for such parameters of the problem are not presented here.
The results of calculations performed using the fully nonlinear dispersion model of \cite {GK_DD_2020a} are very close to that obtained by the Pot-model (Fig.~\ref{Lu}), and this also confirms the validity of the results presented here.

\begin{figure}[h!]
\centering
\includegraphics[width=0.49\textwidth]{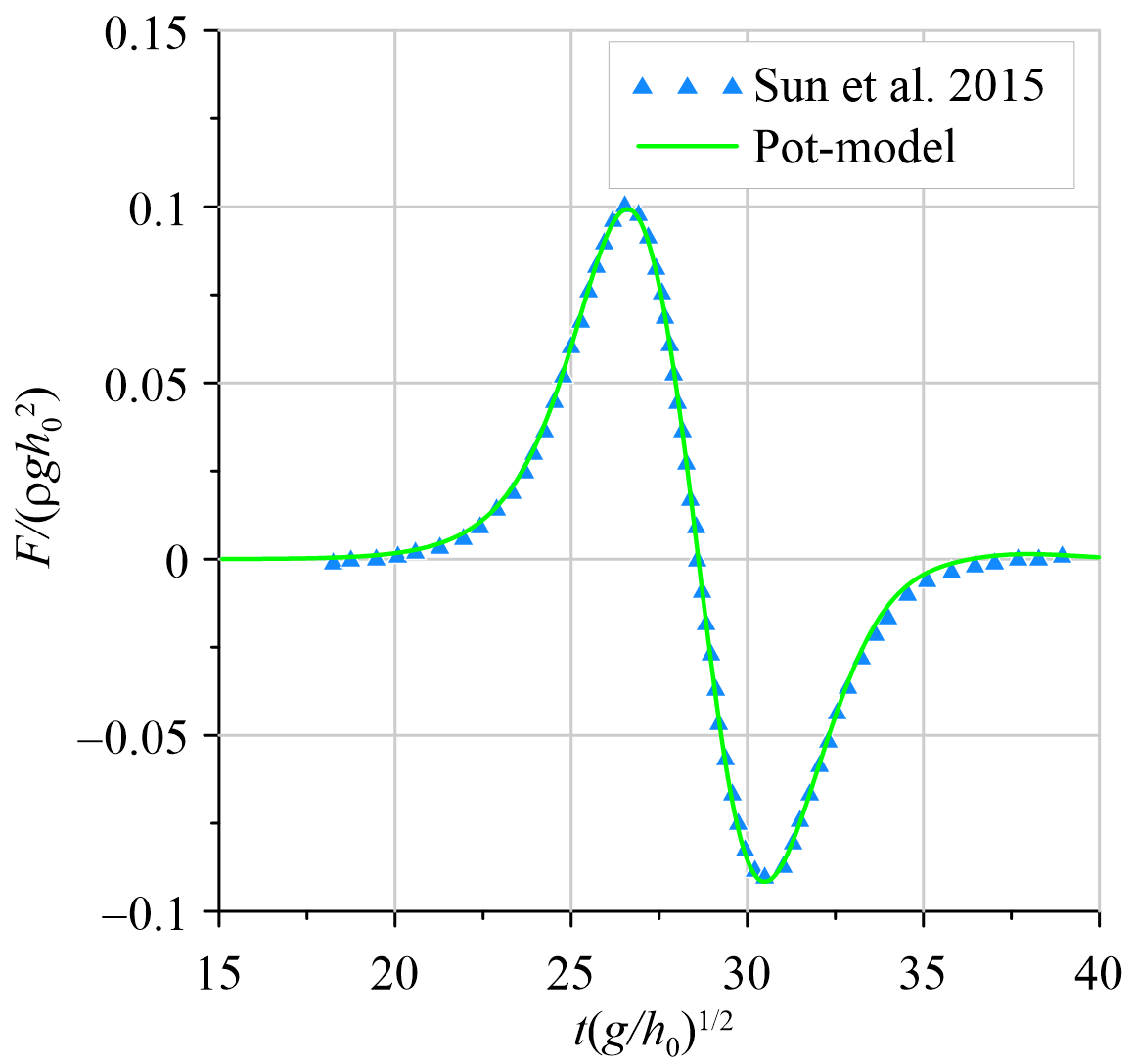}
\\
{\caption{Chronograms of the total horizontal component of the force exerted by the solitary wave~\citep{Sun_etal_2015} of amplitude $a_0/h_0=0.2$ on the structure of length $L/h_0=1$ immersed to the depth of $ d_0 / h_0 = -0.5 $.
\label{Sun_chrono}}}
\end{figure}

\begin{figure}[h!]
\centering
\includegraphics[width=0.49\textwidth]{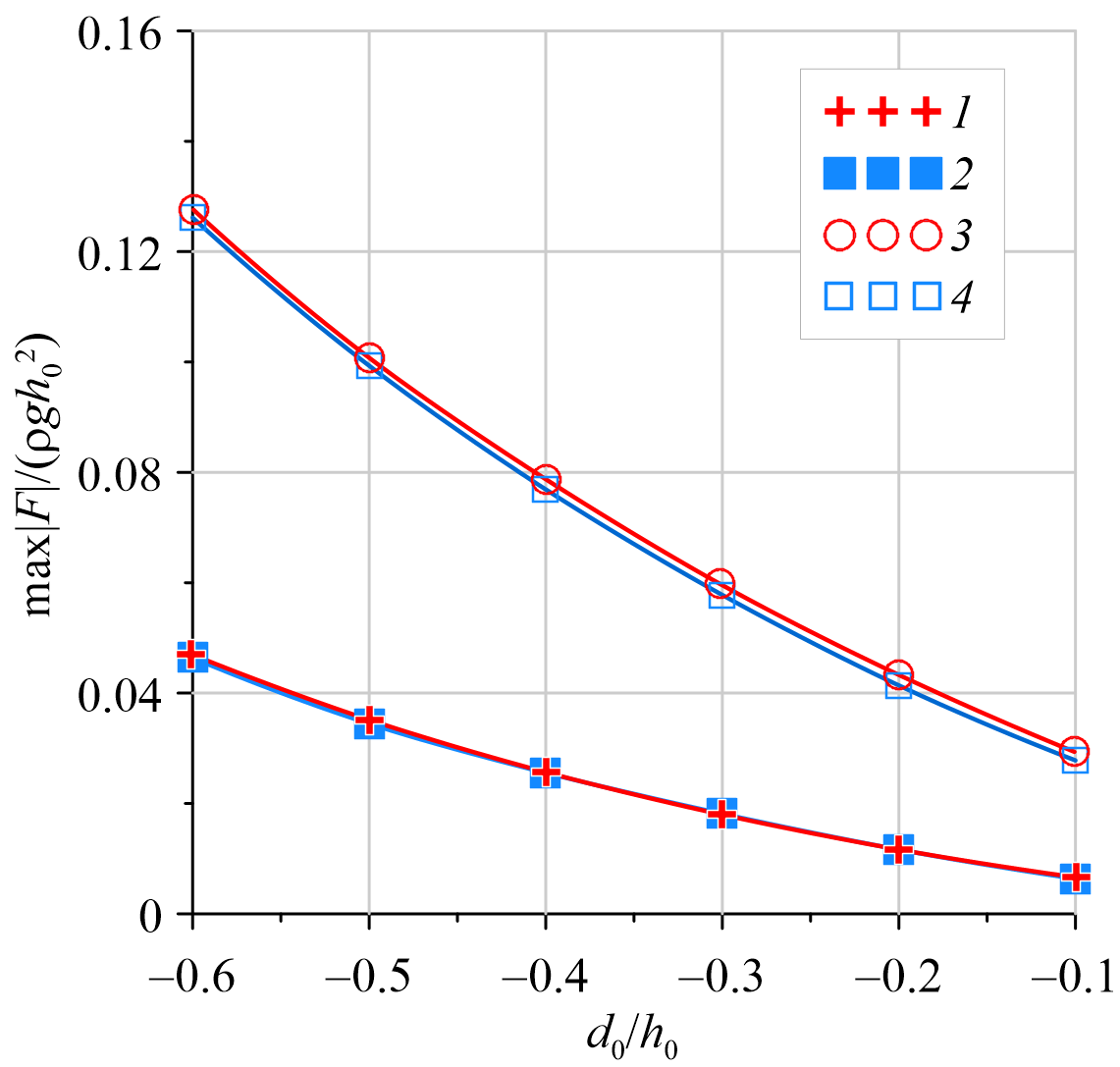}
\\
{\caption{Dependence of the total horizontal component of the force on the structure submergence $ d_0 $ for the solitary wave~\citep{Sun_etal_2015} of amplitude $ a_0 / h_0 = 0.1 $~({\sl 1, 2}) and $ a_0 / h_0 = 0.2 $~({\sl 3, 4}). {\sl 1, 3}, calculation results of  \cite{Sun_etal_2015};  {\sl 2, 4}, Pot-model solutions of the present study.
\label{Sun_vary_d}}}
\end{figure}

\cite{Sun_etal_2015} presented the results of calculations of the wave force acting on a partially immersed structure of length $ L / h_0 = 1 $ and various submergences: from $ d_0 / h_0 = -0.1 $ to $ d_0 / h_0 = -0.6 $. The simulations were performed in the framework of potential flow model using the finite element method. The comparisons of the results of \cite{Sun_etal_2015} (digitaized) with those obtained in the framework of the present Pot-model (see Figs.~\ref{Sun_chrono}, \ref{Sun_vary_d}) demonstrate an excellent agreement. Note that we used  in these computations the same shape of initial solitary wave as in \citep{Sun_etal_2015}. 

\subsection{General characteristics of the long wave impact on a partially immersed and fixed structure}\label{general_view}

\begin{figure}[t!]
\centering
\ \ \includegraphics[width=1.0\textwidth]{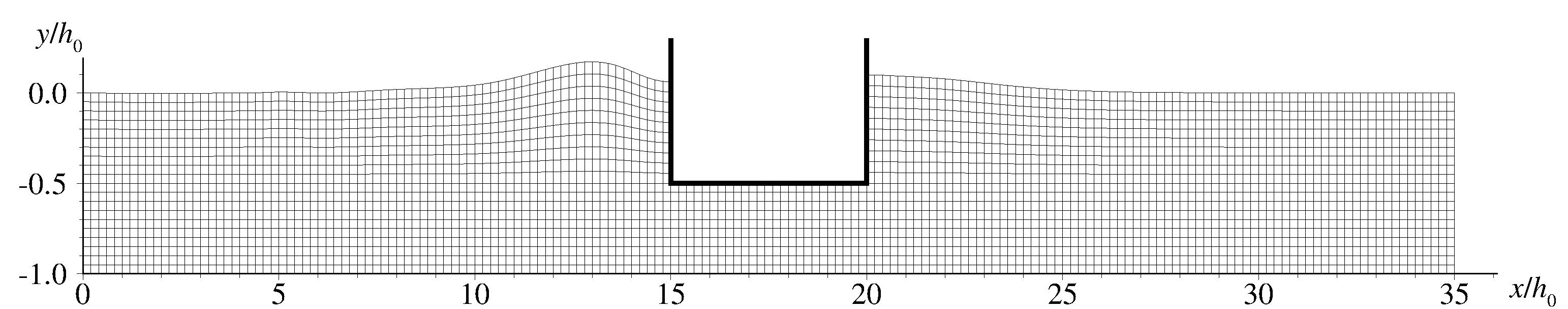}

{\caption{The sample of the computational grid for the problem on interaction of a single wave with a partially immersed structure. $t\sqrt{g/h_0}=9.4$,  $a_0/h_0=0.2$, $\lambda/h_0=10$, $L/h_0=5$, $d_0/h_0=-0.5$, $x_l/h_0=15$, $x_0=x_l/2$. Each second grid coordinate line is shown.}
\label{Grid_Pot}}
\end{figure}

The computations were performed with the  grid with moving nodes (Fig.~\ref{Grid_Pot}). In the simplified 2D formulation, the resource intensity of the calculations is relatively low, so here we used a rather fine grid uniform along the $Ox$ axis with a step $\Delta x/h_0=0.1$. 
For very long waves (single waves with length $\lambda/h_0=50$ and $\lambda/h_0=250$), the irregular grid is used for $Ox$ axis to reduce the computational cost. Namely, to the left of the structure, the grid step increased exponentially with increasing the distance from the body. This approach allows to use a small number of grid nodes ($N_1=400$) and at the same time maintain high accuracy of the calculations near the structure where the most important changes in the flow occur. The advantages of this approach in 3D-calculations are even more pronounced.

The physical domain was divided along the $Oy$ axis into $N_2=40$ layers, and their number  above the lower edge of the body ($N_2-j_b$) was equal to $\max(N_2|d_0|/h_0,8)$, i.e. it had to be at least eight. 
 This grid resolution proved to be quite sufficient, since further refinement changes the maximum wave force by less than 1\% for each set of the problem parameters considered below.

\begin{figure}[h!]
\centering
\parbox[t]{0.49\textwidth}{\centering {\it a}} \hfill \parbox[t]{0.49\textwidth}{\centering {\it b}}\\
\includegraphics[width=0.49\textwidth]{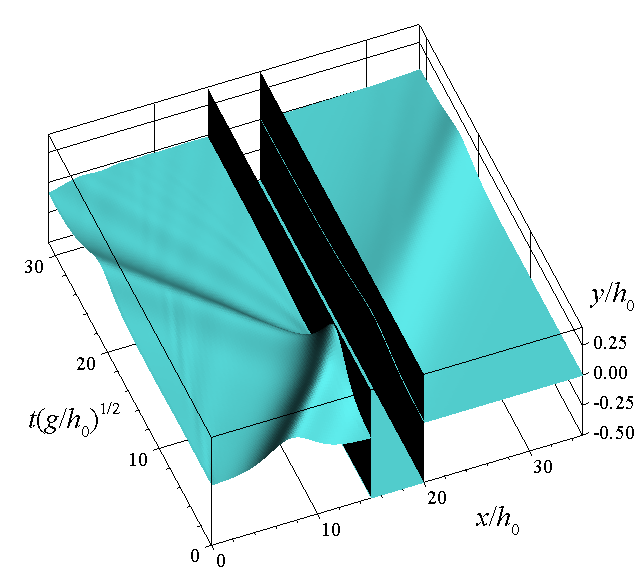}\hfill
\includegraphics[width=0.49\textwidth]{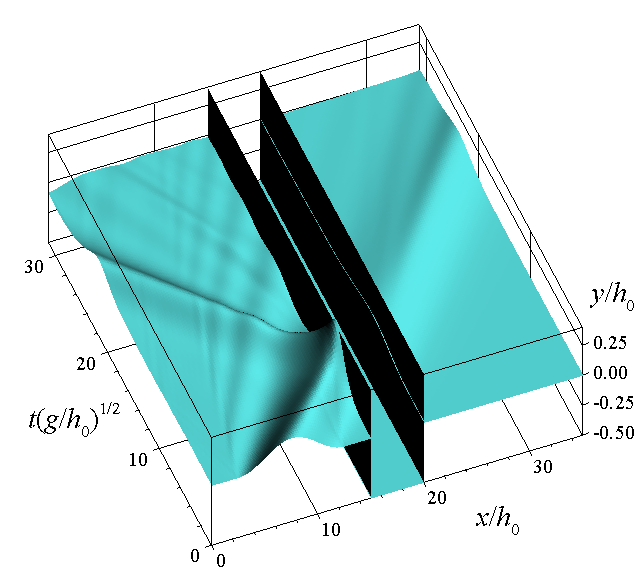}

{\caption{Dynamics of the free surface $y=\eta(x,t)$ when the solitary ({\it a}) and single ({\it b}) (with the length $ \lambda / h_0 = 10 $) waves interact with a partially immersed and fixed structure.
$a_0/h_0=0.2$, $L/h_0=5$, $d_0/h_0=-0.5$, $x_l/h_0=15$, $x_0=x_l/2$.}
\label{Surface_with_refl_conditions}}
\end{figure}

A general views of the interaction of long surface wave with a partially immersed and fixed structure are presented at Figure~\ref{Surface_with_refl_conditions}. It shows the dynamics of the free surface computed for the solitary (Fig.~\ref{Surface_with_refl_conditions},~{\it a}) and single (Fig.~\ref{Surface_with_refl_conditions},~{\it b}) waves.
The reflected wave resulting from such interaction has an amplitude less than the initial $a_0$.
It should also be noted the formation after the reflection  of the  moving  depression and dispersive wave train following it. 
A new solitary wave with, naturally, an amplitude smaller than $a_0$ is formed from the wave transmitted beyond the structure. 
These characteristics are common for the two types of waves. In the case of a single wave the picture is complicated by the interaction with the structure of the dispersive wave train formed during the above-mentioned soliton formation.

The computational experiments were performed for study the dependence of the force on such characteristics as the amplitude and length of the wave, the submergence and length of the partially immersed structure. Only one of the parameters of the problem was varied during each calculation, and the values of all the others did not change.
The following set of problem parameters was considered as ``basic'': the wave amplitude $a_0/h_0 = 0.1$, its length $\lambda / h_0 = 10 $, the length of the structure $ L / h_0 = 5 $, its submergence $ d_0 / h_0 = -0.5 $.
At the same time, we consider longer waves, $\lambda / h_0 = 50 $ and $\lambda / h_0 = 250 $, which are closer to tsunami problems.
The wave was located at the initial moment of time $ t = 0 $ to the left of the structure and close to it: $ x_l = \lambda $, $ x_0 = \lambda / 2 $. The length of the computational domain was chosen equal to $ l = x_r + 10h_0 $ for all numerical experiments. 
For shorter waves (solitary or single waves with length $\lambda/h_0=10$),  the reflected wave does not have time to influence the estimation of the main component of the force action. For relatively long waves (single waves with length $\lambda/h_0=50$ and $\lambda/h_0=250$), it was checked that the boundary conditions used by the authors lead to a small, no more than 1\%, perturbation of the calculated force characteristic.

\subsection{Dependence of the wave force on the problem parameters}\label{main_results}

The dependences of the wave force on the submergence of the structure, its length, and the amplitude of the incident wave are discussed in this subsection.  Furthermore, the results obtained for single waves of various lengths and for the solitary wave are compared. 

\begin{figure}[h!]
\centering
\includegraphics[width=0.49\textwidth]{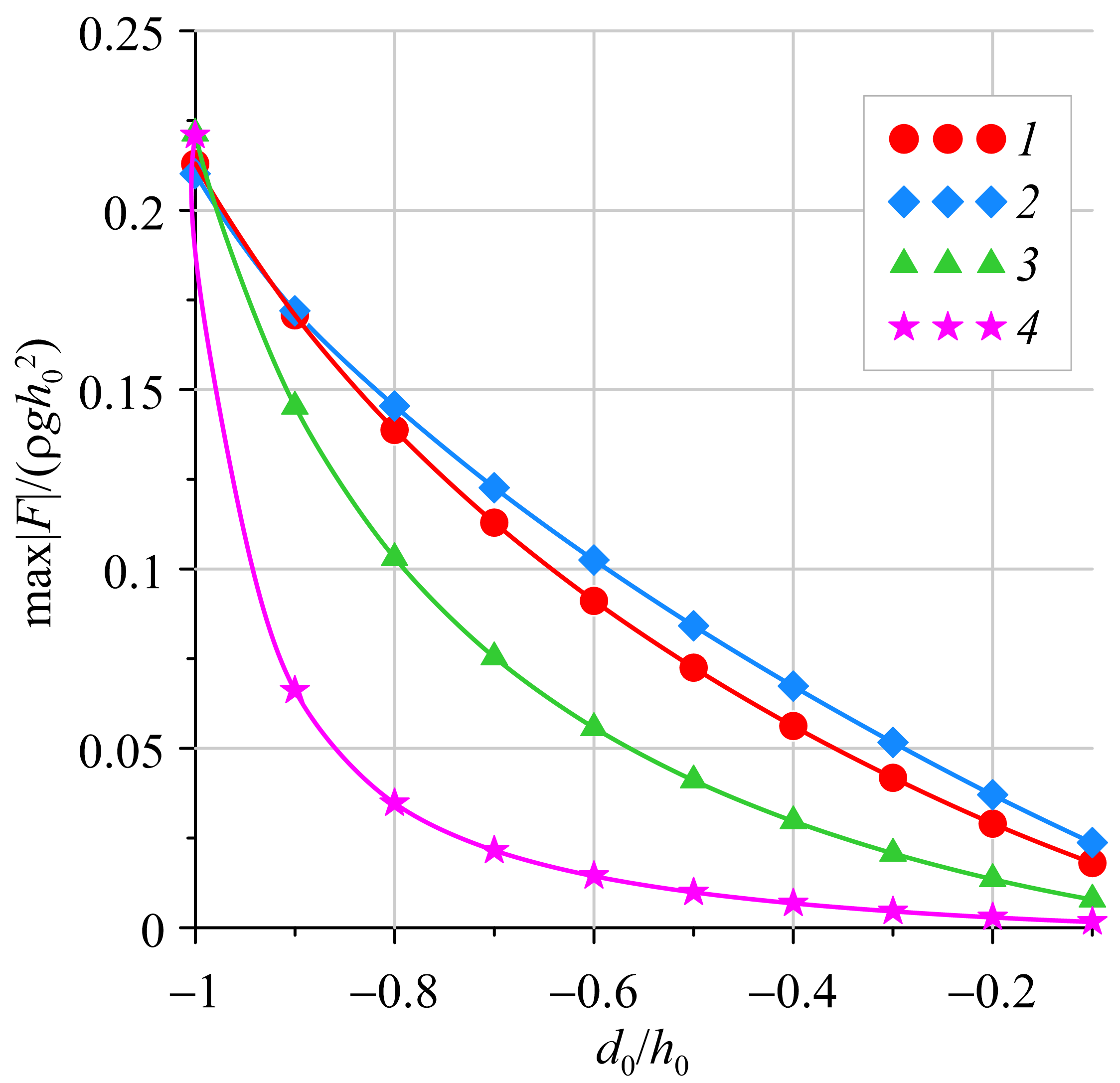}
\\
{\caption{Dependence of the total horizontal component of the force on the structure submergence $ d_0 $ for the solitary wave~({\sl 1}) and single wave of length~$\lambda/h_0=10$~({\sl 2}),~$50$~({\sl 3}) and $250$~({\sl 4}).
$a_0/h_0=0.1$, $L/h_0=5$.}
\label{F_d}}
\end{figure}

The computed results allow to determine the dependence of the total horizontal component of the wave force on the submergence of the structure $d_0$ (see Fig.~\ref{F_d}), where the submergence $d_0/h_0=-1$ corresponds to the problem  of the wave interaction with a vertical wall. 
Note that in this figure and most of the ones below, we consider the maximum force value over the entire computation time.
These results show that the total horizontal wave force increases with increasing submergence of the structure, which is consistent with the conclusions of \cite{Chen_Wang_2019,Lu_Wang_2015, Sun_etal_2015}, and can be explained by the fact that with increasing submergence, the structure becomes more significant obstacle for the wave.  The wave impact on the front side of the structure increases, and it decreases on the back side \citep{Sun_etal_2015}.
The same behaviour was observed by \citep{Mavrakos_1985} in the study of wave loads  on a stationary floating bottomless cylindrical body.
An indirect confirmation of these conclusions is the results of \cite{Losada} on the impact of waves on thin impenetrable barriers, where, in particular, it is shown that the horizontal component of the force on a partially immersed barrier increases with increasing submergence. 
Analytical calculations \citep{Fang_Guo_2019} of the wave force acting on a partially immersed box when subjected to a focused wave group attack showed that  the horizontal wave force increases with the increase of submergence, if this force is written in the dimensionless form of this study, $\displaystyle\frac{F}{\rho g h_0^2}$.

The force characteristics calculated for the solitary and single ($\lambda/h_0=10$) waves (curves {\sl 1} and {\sl 2} in Fig.~\ref{F_d}) are very close. The force induced by the solitary wave is somewhat smaller at small submergences.
Comparisons of the curves {\sl 2}, {\sl 3} and {\sl 4} in this figure shows that 
when the length of the single wave increases then
the force on the partially immersed structure decreases and the force on the vertical wall (at $ d_0 / h_0 = -1 $) increases. Also, the force characteristic of the longer single waves has the greater curvature.

\begin{figure}[h!]
\centering
\includegraphics[width=0.49\textwidth]{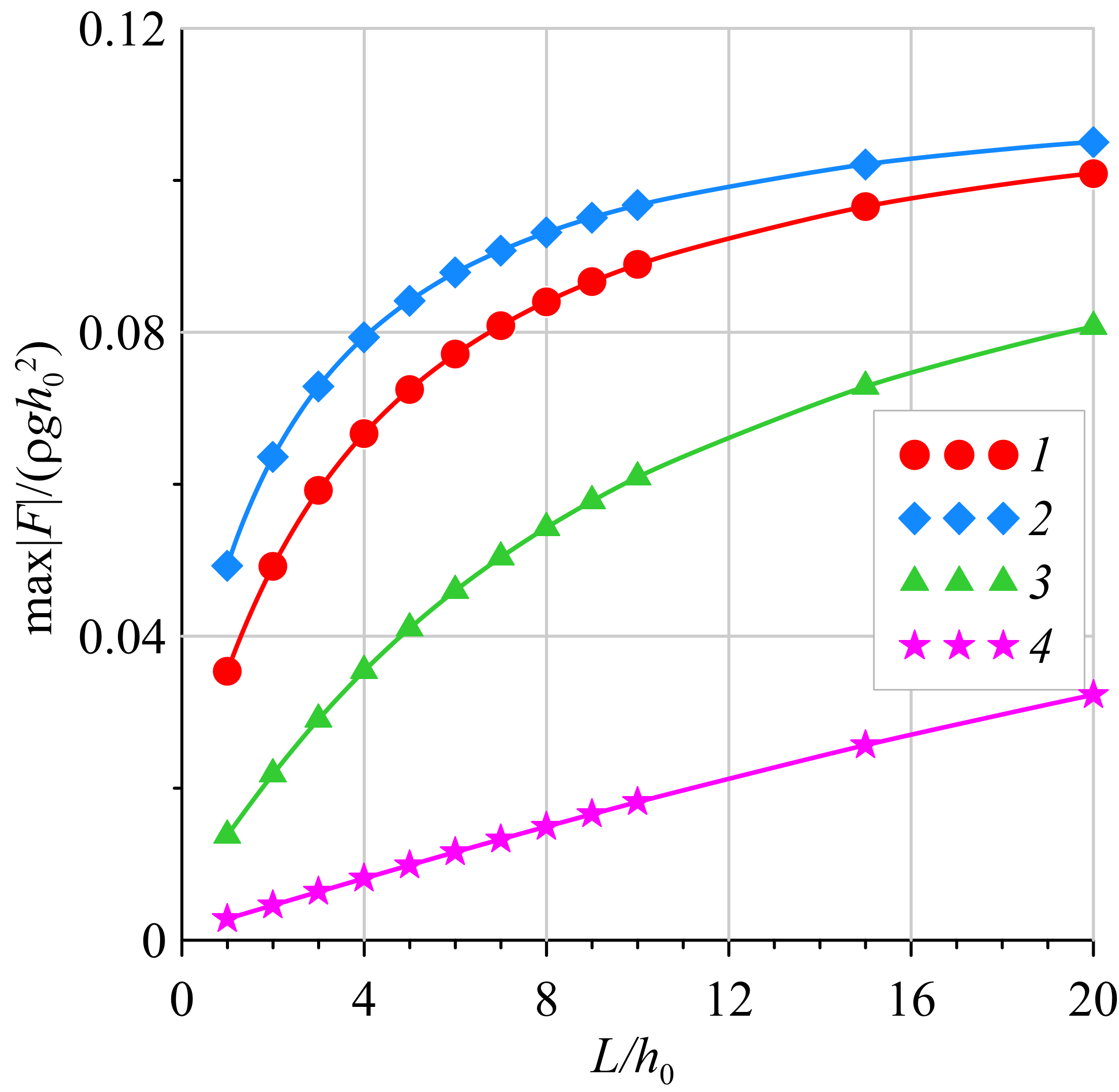}
\\
{\caption{Dependence of the total horizontal component of the force on the structure length for the solitary wave~({\sl 1}) and single wave of length $\lambda/h_0=10$~({\sl 2}),~$50$~({\sl 3}) and $250$~({\sl 4}).
$a_0/h_0=0.1$, $d_0/h_0=-0.5$. }
\label{F_L}}
\end{figure}

The dependence of the total horizontal component of the force on the structure length (Fig.~\ref{F_L}) shows that this force increases with increasing the length.
The solitary wave has a slightly smaller effect than the shorter single wave ($\lambda / h_0 = 10 $). The corresponding dependences {\sl 1} and {\sl 2} are also close  in general. 
When the incident wave length increases, the force decreases for each structure length, and the curvature of the corresponding curve also decreases. 
It is worth noting that for very long waves ($\lambda / h_0 = 250 $) this dependence is almost linear.

\begin{figure}[h!]
\centering
\includegraphics[width=0.49\textwidth]{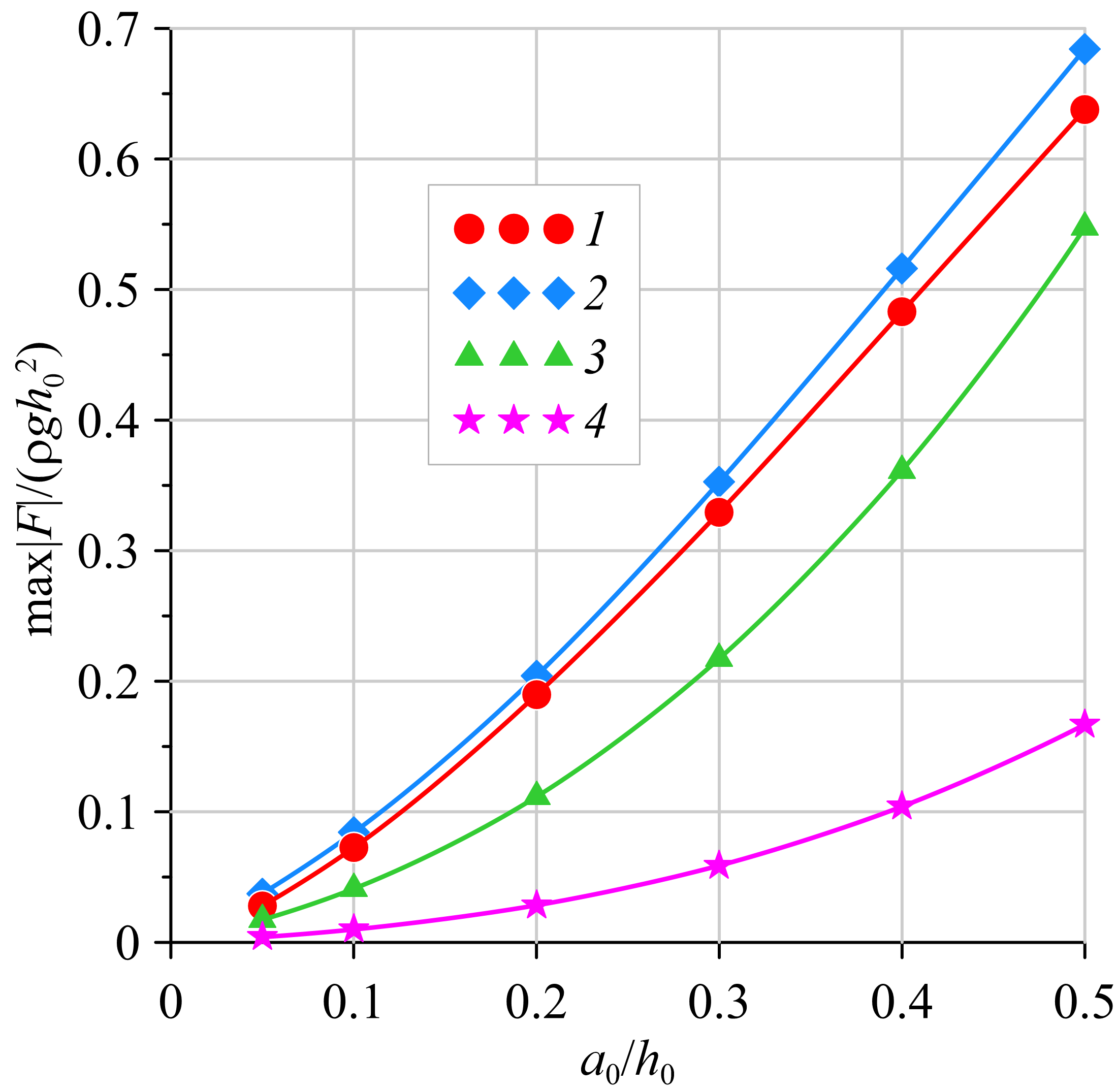}
\\
{\caption{Dependence of the total horizontal component of the force on the amplitude $a_0$ for the solitary wave~({\sl 1}) and for single waves with the length $\lambda/h_0=10$~({\sl 2}),~$50$~({\sl 3}) and $250$~({\sl 4}).
$d/h_0=-0.5$, $L/h_0=5$.}
\label{F_a}}
\end{figure}

\begin{figure}[h!]
\centering
\includegraphics[width=0.49\textwidth]{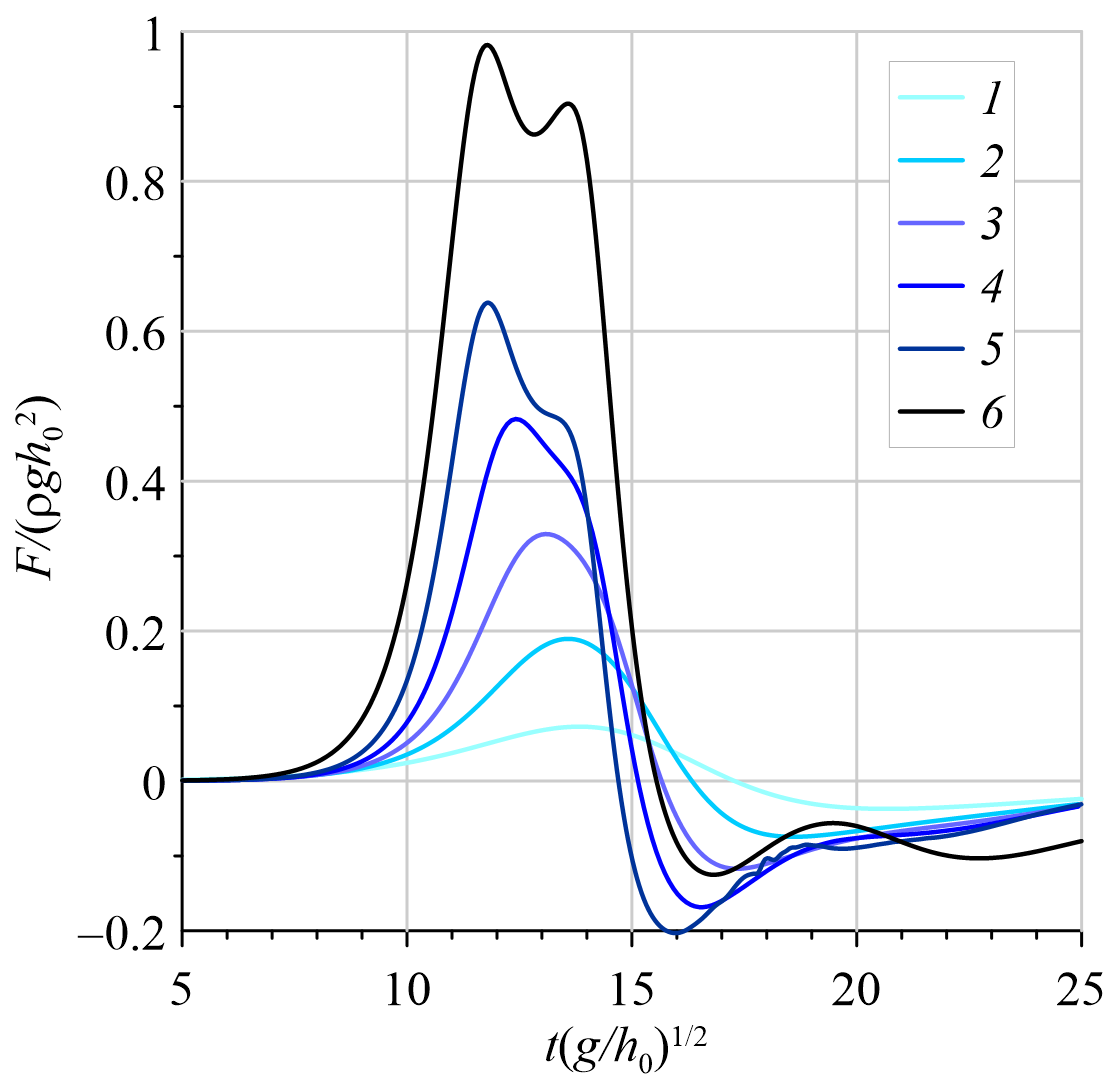}
\\
{\caption{Chronograms of the total horizontal component of the force acting on the partially immersed structure for the solitary wave of amplitude  $a_0/h_0=0.1$ ({\sl 1}), $0.2$ ({\sl 2}), $0.3$ ({\sl 3}), $0.4$ ({\sl 4}) and $0.5$~({\sl 5, 6}) on the partially immersed structure.
$d_0/h_0=-0.5$ ({\sl 1--5}), $-0.9$ ({\sl 6}). $L/h_0=5$.
} 
\label{F_a_chr}}
\end{figure}

The dependence of the force on the amplitude of  incident wave $a_0$ is shown in Figure~\ref{F_a}. 
Its behaviour does not contradict the expectation that the force will increase when the  amplitude increases. 
The dependences calculated for the solitary wave (curve {\sl 1}) and for single wave with the length $ \lambda / h_0 = 10 $ (curve {\sl 2}) are very close, as in the analysis of the dependence of the force characteristic on the structure length. In this case, the solitary wave for each amplitude has a slightly smaller effect. 
For longer waves, $ \lambda / h_0 = 50 $ and $250$, the force is less, and the curvature of the corresponding curves ({\sl 3, 4}) is greater.
It should also be noted that in the force chronograms (see~Fig.~\ref{F_a_chr}) for the waves with an amplitude of $a_0/h_0 \ge 0.5 $, a second local maximum appears. This effect occurs only at large submergences of the structure (curve {\sl 6 } for $ d_0 / h_0 = -0.9 $) and is also known for the interaction of a solitary wave with a fixed vertical wall \citep[e.g.]{Cooker_1997, Madsen_Bingham_Liu_2002}.

\begin{figure}[h!]
\centering
\includegraphics[width=0.49\textwidth]{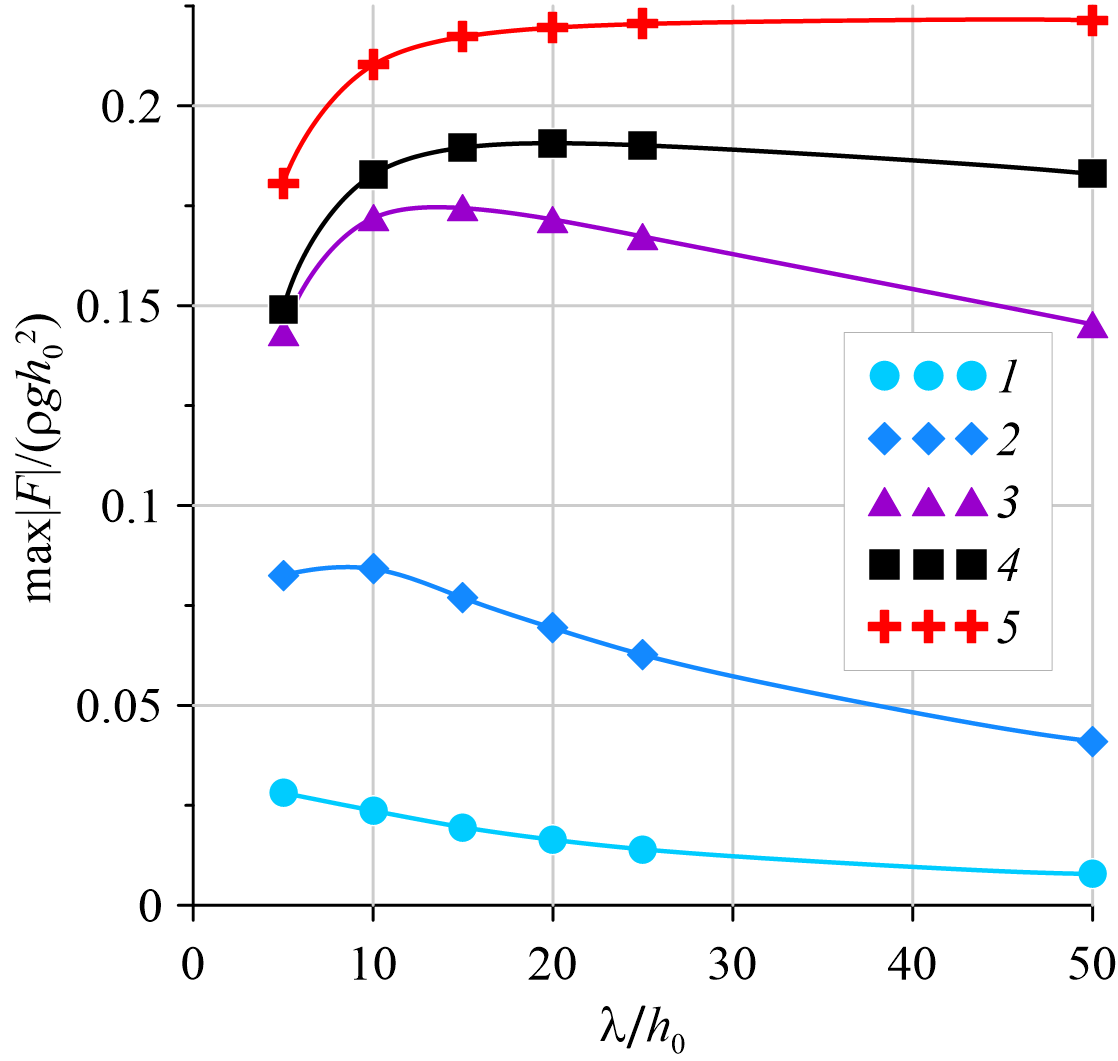}\hfill
\\
{\caption{Dependence of the total horizontal component of the force on the length $\lambda$ of the single wave  of amplitude $a_0/h_0=0.1$: \ {\sl 1}--{\sl 4}, the wave impact on the partially immersed structure with $d_0/h_0=-0.1$, $L/h_0=5$~({\sl 1}); \ $d_0/h_0=-0.5$, $L/h_0=5$~({\sl 2}); \ $d_0/h_0=-0.9$, $L/h_0=5$~({\sl 3}); \  $d_0/h_0=-0.9$, $L/h_0=20$~({\sl 4}); the wave impact on the vertical wall ({\sl 5}).
}
\label{F_Lw}}
\end{figure}

The dependences of the force exerted by the single wave on its length  are presented on Figure~\ref{F_Lw}. 
Several options of structure submergence  are considered in this paper for the ``basic'' length.
The option with the very submerged and very wide structure was additionally considered, which in some way brings the problem closer to modeling the interaction of a wave with a vertical wall. The results of wave impact directly on the vertical wall  ($ d_0 / h_0 = -1 $) are also presented. 

The wave force increases monotonically with increasing wavelength when the wave impacts on the vertical wall  (curve {\sl 5} in Fig.~\ref{F_Lw}). In contrast, the wave force decreases monotonically for the considered wavelengths at small submergence and small length of the structure (curve {\sl 1} in Fig.~\ref{F_Lw}). Long waves hardly notice the object in this case.

\begin{figure}[h!]
\centering
\includegraphics[width=0.49\textwidth]{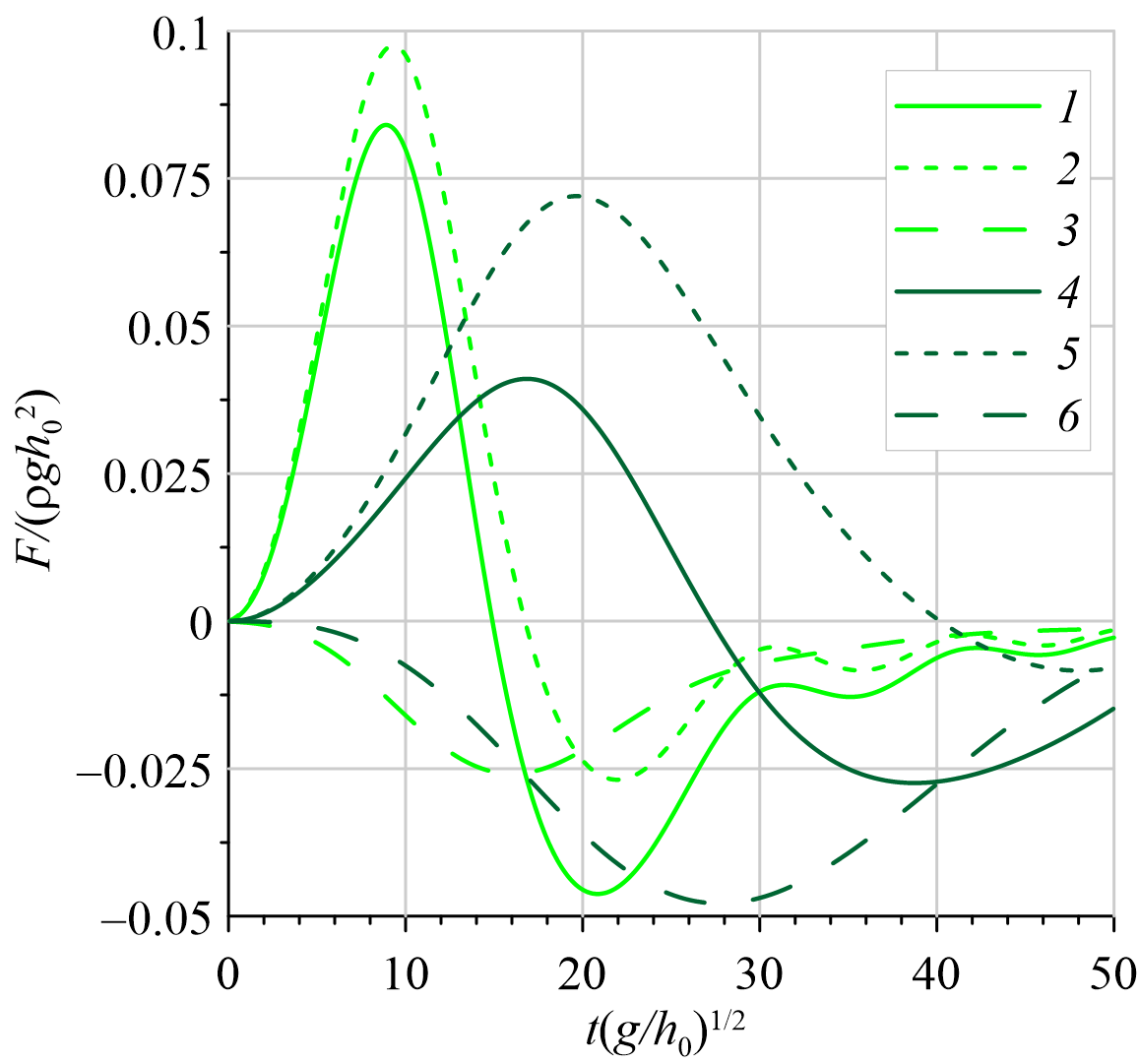}
\\
{\caption{Chronograms of the total horizontal component of the force for the single wave on the partially immersed structure: the total force ({\sl 1, 4}) and forces on the left ({\sl 2, 5}) and right ({\sl 3, 6}) faces.
$\lambda/h_0=10$~({\sl 1, 2, 3}) and $\lambda/h_0=50$~({\sl 4, 5, 6}), 
$a_0/h_0=0.1$, $d_0/h_0=-0.5$, $L/h_0=5$. }
\label{F_Lw_chr}}
\end{figure}

The forces acting in opposite directions on the front and back faces of the structure change as the wavelength increases (see Fig.~\ref {F_Lw_chr}), and the maximum of
the total force (taking into account the direction of action) decreases in absolute value.
Some indirect confirmation of the revealed pattern can be provided by the theoretical results of  \cite{Lee_Chwang_2000} on the wave force on protective screens of various porosities. An analogy with the problem considered here arises for sufficiently long waves which allows to identify the immersed structure with a thin screen submerged in the water to a certain depth. The aforementioned theoretical results indicate that the horizontal component of the force increases rapidly with decreasing incident wavelength, which is consistent with the intuitive notion that long waves are weakly reflected from the screen, while short ones are reflected significantly.

With a decrease in the thickness of the water layer under the structure, this trend weakens and the dependence naturally approaches that noted for the vertical wall. 
The dependence of the force on the wavelength is non-monotonic in the cases when the body is deeply submerged (curves {\sl 2--4} in Fig.~\ref{F_Lw}), i.e.  the maximum appears for a certain wavelength. 
Note that the corresponding wavelength increases with an increase in the submergence of the structure and its length. 
The non-monotonic character of the dependence of the force on the wavelength was also established in the study of \cite{Lee_Chwang_2000} for a completely submerged in water and fixed at the bottom thin screen. 
This effect was confirmed in the numerical results presented here for a partially immersed structure of finite length. The non-monotonic dependence of the horizontal component of the wave force was also noticed by \cite {Chen_2019} when studying the wave force acting on the first of the three successively installed bodies of a floating module.

\section{Conclusions}
In this study, the dependence of the horizontal component of the force exerted by  long surface waves on a partially immersed and fixed structure on the wave length and amplitude, submergence and length of the body is investigated. The solitary wave and single waves of finite length were used as the incoming waves. It is shown that with an appropriate choice of the length of the single wave, the computed force exerted by the single and solitary waves are quite close.

It is established that the force impact increases with increasing the submergence and length of the structure, as well as with increasing the amplitude of the incoming wave.
It is noteworthy that at large wave amplitudes and structure submergences, a second local maximum appears in the force action chronograms. It is similar to the problem of the interaction of a wave with a vertical wall.
	
One of the most important parameters of the problem is the wavelength: the force decreases monotonically at small submergences of the structure, when wavelength increases, while the force acting on the vertical wall monotonically increases. The dependence of the force on the wavelength is non-monotonic in other cases, a force maximum is detected at a certain wavelength. This ``resonant'' wavelength increases with increasing submergence and length of the body.

The comparisons of the results obtained within the framework of the potential flow model with the data of small-scale laboratory experiments and the results of calculations of the other authors in the problem of the interaction of a wave with a fixed vertical wall confirm the validity of the results presented  in the paper. The substantial  deficiency of publications devoted to the physical modeling of wave interaction with a partially immersed and fixed body, contained  the measurement data of the force acting on it, should be noted.

Further development of the research we expect in the direction of expanding the spectrum (hierarchy) of mathematical models and algorithms. The practical significance of the hierarchical approach, where problems are studied using interacting mathematical models of wave hydrodynamics of various approximations, consists in increasing of the degree of reliability of numerical results and in substantiated determination of the ranges of applicability of low-level models. The use of less costly algorithms so long as they provide the required accuracy and take into account key physical effects, significantly saves  the computational resources especially when performing multivariant  calculations. We also plan to extend this approach for a three-dimensional case and to develop it for the formulation of the problems in curvilinear coordinates for partially submerged bodies of complex shape.

\section*{Acknowledgements} 
We thank the anonymous Reviewers for their very helpful comments on this manuscript.
The results were obtained within the framework of the theme No. 0316-2019-0001 ``Development and research of new elements of computational technology for solving fundamental and applied problems of aero-, hydro- and wave dynamics'' of the state task of the Federal Research Center for Information and Computational Technologies.


\bibliographystyle{cas-model2-names}

\bibliography{cas-refs}

\begin{thebibliography}{32}
\expandafter\ifx\csname natexlab\endcsname\relax\def\natexlab#1{#1}\fi
\providecommand{\url}[1]{\texttt{#1}}
\providecommand{\href}[2]{#2}
\providecommand{\path}[1]{#1}
\providecommand{\DOIprefix}{doi:}
\providecommand{\ArXivprefix}{arXiv:}
\providecommand{\URLprefix}{URL: }
\providecommand{\Pubmedprefix}{pmid:}
\providecommand{\doi}[1]{\href{http://dx.doi.org/#1}{\path{#1}}}
\providecommand{\Pubmed}[1]{\href{pmid:#1}{\path{#1}}}
\providecommand{\bibinfo}[2]{#2}
\ifx\xfnm\relax \def\xfnm[#1]{\unskip,\space#1}\fi
\bibitem[{Bona et~al.(2005)Bona, Colin and Lannes}]{Bona_2005}
\bibinfo{author}{Bona, J.L.}, \bibinfo{author}{Colin, T.},
  \bibinfo{author}{Lannes, D.}, \bibinfo{year}{2005}.
\newblock \bibinfo{title}{Long wave approximations for water waves}.
\newblock \bibinfo{journal}{Arch. Ration. Mech. An.} \bibinfo{volume}{178},
  \bibinfo{pages}{373--410}.
\newblock \DOIprefix\doi{https://doi.org/10.1007/s00205-005-0378-1}.
\bibitem[{Bresch et~al.(2019)Bresch, Lannes and Metivier}]{Bresch_Lannes_2019}
\bibinfo{author}{Bresch, D.}, \bibinfo{author}{Lannes, D.},
  \bibinfo{author}{Metivier, G.}, \bibinfo{year}{2019}.
\newblock \bibinfo{title}{Waves interacting with a partially immersed obstacle
  in the {B}oussinesq regime} \URLprefix
  \url{http://arxiv.org/abs/1902.04837v1},
  \href{http://arxiv.org/abs/http://arxiv.org/abs/1902.04837v1}{\tt
  arXiv:http://arxiv.org/abs/1902.04837v1}.
\bibitem[{Chang(2017)}]{Chang_2017}
\bibinfo{author}{Chang, C.H.}, \bibinfo{year}{2017}.
\newblock \bibinfo{title}{Study of a solitary wave interacting with a surface
  piercing square cylinder using a three-dimensional fully nonlinear model with
  grid-refinement technique on surface layers}.
\newblock \bibinfo{journal}{J. Mar. Eng. Technol.} \bibinfo{volume}{16 (1)},
  \bibinfo{pages}{22--36}.
\newblock \DOIprefix\doi{https://doi.org/10.1080/20464177.2016.1277605}.
\bibitem[{Chang et~al.(2017)Chang, Wang and Hseih}]{Chang_Wang_Hseih_2017}
\bibinfo{author}{Chang, C.H.}, \bibinfo{author}{Wang, K.H.},
  \bibinfo{author}{Hseih, P.C.}, \bibinfo{year}{2017}.
\newblock \bibinfo{title}{Fully nonlinear model for simulating solitary waves
  propagating through a partially immersed rectangular structure}.
\newblock \bibinfo{journal}{J. Coastal Res.} \bibinfo{volume}{33 (6)},
  \bibinfo{pages}{1487--1497}.
\newblock \DOIprefix\doi{https://doi.org/10.2112/JCOASTRES-D-16-00061.1}.
\bibitem[{Chen et~al.(2019a)Chen, He, Bingham and Shao}]{Chen_2019}
\bibinfo{author}{Chen, L.}, \bibinfo{author}{He, G.}, \bibinfo{author}{Bingham,
  H.B.}, \bibinfo{author}{Shao, Y.}, \bibinfo{year}{2019}a.
\newblock \bibinfo{title}{Gap resonance of fixed floating multi caissons}, in:
  \bibinfo{booktitle}{Proc. of the {ASME} 2019 38th {I}nternational Conference
  on Ocean, Offshore and Arctic Engineering ({V}ol. 7{A}: Ocean Engineering)},
  \bibinfo{publisher}{American Society of Mechanical Engineers}. pp.
  \bibinfo{pages}{[OMAE2019--96383]}.
\newblock \DOIprefix\doi{https://doi.org/10.1115/OMAE2019-96383}.
\bibitem[{Chen and Wang(2019)}]{Chen_Wang_2019}
\bibinfo{author}{Chen, L.}, \bibinfo{author}{Wang, K.H.}, \bibinfo{year}{2019}.
\newblock \bibinfo{title}{Experiments and computations of solitary wave
  interaction with fixed, partially submerged, vertical cylinders}.
\newblock \bibinfo{journal}{J. Ocean Eng. Mar. Energy} \bibinfo{volume}{5},
  \bibinfo{pages}{189–204}.
\newblock \DOIprefix\doi{https://doi.org/10.1007/s40722-019-00137-8}.
\bibitem[{Chen et~al.(2019b)Chen, Li, Hsu and Hwung}]{Chen_et_al_2019}
\bibinfo{author}{Chen, Y.Y.}, \bibinfo{author}{Li, Y.J.}, \bibinfo{author}{Hsu,
  H.C.}, \bibinfo{author}{Hwung, H.H.}, \bibinfo{year}{2019}b.
\newblock \bibinfo{title}{The pressure distribution beneath a solitary wave
  reflecting on a vertical wall}.
\newblock \bibinfo{journal}{Eur. J. Mech. B-Fluid} \bibinfo{volume}{76},
  \bibinfo{pages}{66--72}.
\newblock \DOIprefix\doi{https://doi.org/10.1016/j.euromechflu.2019.01.010}.
\bibitem[{Cooker et~al.(1997)Cooker, Weidman and Bale}]{Cooker_1997}
\bibinfo{author}{Cooker, M.J.}, \bibinfo{author}{Weidman, P.D.},
  \bibinfo{author}{Bale, D.S.}, \bibinfo{year}{1997}.
\newblock \bibinfo{title}{Reflection of a high-amplitude solitary wave at a
  vertical wall}.
\newblock \bibinfo{journal}{J. Fluid Mech.} \bibinfo{volume}{342},
  \bibinfo{pages}{141--158}.
\newblock \DOIprefix\doi{https://doi.org/10.1017/S002211209700551X}.
\bibitem[{Engsig-Karup et~al.(2017)Engsig-Karup, Monteserin and
  Eskilsson}]{Engsig_2017}
\bibinfo{author}{Engsig-Karup, A.P.}, \bibinfo{author}{Monteserin, C.},
  \bibinfo{author}{Eskilsson, C.}, \bibinfo{year}{2017}.
\newblock \bibinfo{title}{A stabilised nodal spectral element method for fully
  nonlinear water waves. {P}art 2: Wave-body interaction} \URLprefix
  \url{http://arxiv.org/abs/1703.09697v1},
  \href{http://arxiv.org/abs/http://arxiv.org/abs/1703.09697v1}{\tt
  arXiv:http://arxiv.org/abs/1703.09697v1}.
\bibitem[{Fang and Guo(2019)}]{Fang_Guo_2019}
\bibinfo{author}{Fang, Q.}, \bibinfo{author}{Guo, A.}, \bibinfo{year}{2019}.
\newblock \bibinfo{title}{Analytical and experimental study of focused wave
  action on a partially immersed box}.
\newblock \bibinfo{journal}{Math. Probl. Eng.} \bibinfo{volume}{2019}.
\newblock \DOIprefix\doi{https://doi.org/10.1155/2019/9850302}.
  \bibinfo{note}{article ID 9850302, 15 p.}
\bibitem[{Kamynin et~al.(2010)Kamynin, Maximov, Nudner, Semenov and
  Khakimzyanov}]{Kamynin_2010}
\bibinfo{author}{Kamynin, E.Y.}, \bibinfo{author}{Maximov, V.V.},
  \bibinfo{author}{Nudner, I.S.}, \bibinfo{author}{Semenov, K.K.},
  \bibinfo{author}{Khakimzyanov, G.S.}, \bibinfo{year}{2010}.
\newblock \bibinfo{title}{Study of interaction of the solitary wave with a
  partially submerged stationary construction}.
\newblock \bibinfo{journal}{Fundamental and Applied Hydrophysics}
  \bibinfo{volume}{4 (10)}, \bibinfo{pages}{39--54}.
\newblock \bibinfo{note}{(In Russ.)}.
\bibitem[{Khakimzyanov and Dutykh(2018)}]{GK_DD_2018e}
\bibinfo{author}{Khakimzyanov, G.}, \bibinfo{author}{Dutykh, D.},
  \bibinfo{year}{2018}.
\newblock \bibinfo{title}{Numerical modelling of surface water wave interaction
  with a moving wall}.
\newblock \bibinfo{journal}{Commun. Comput. Phys.} \bibinfo{volume}{23 (5)},
  \bibinfo{pages}{1289--1354}.
\newblock \DOIprefix\doi{10.4208/cicp.OA-2017-0110}.
\bibitem[{Khakimzyanov and Dutykh(2020)}]{GK_DD_2020a}
\bibinfo{author}{Khakimzyanov, G.}, \bibinfo{author}{Dutykh, D.},
  \bibinfo{year}{2020}.
\newblock \bibinfo{title}{Long wave interaction with a partially immersed body.
  {Part~I: M}athematical models}.
\newblock \bibinfo{journal}{Commun. Comput. Phys.} \bibinfo{volume}{27 (2)},
  \bibinfo{pages}{321--378}.
\newblock \DOIprefix\doi{10.4208/cicp.OA-2018-0294}.
\bibitem[{Khakimzyanov(2002)}]{Khakimzyanov_2002}
\bibinfo{author}{Khakimzyanov, G.S.}, \bibinfo{year}{2002}.
\newblock \bibinfo{title}{Numerical simulation of the interaction of a solitary
  wave with a partially immersed body}.
\newblock \bibinfo{journal}{Russ. J. Numer. Anal. M.} \bibinfo{volume}{17 (2)},
  \bibinfo{pages}{145–158}.
\newblock \DOIprefix\doi{https://doi.org/10.1515/rnam-2002-0204}.
\bibitem[{Khakimzyanov et~al.(2001)Khakimzyanov, Shokin, Barakhnin and
  Shokina}]{Khakimzyanov_2001}
\bibinfo{author}{Khakimzyanov, G.S.}, \bibinfo{author}{Shokin, Y.I.},
  \bibinfo{author}{Barakhnin, V.B.}, \bibinfo{author}{Shokina, N.Y.},
  \bibinfo{year}{2001}.
\newblock \bibinfo{title}{Numerical Simulation of Fluid Flows with Surface
  Waves}.
\newblock \bibinfo{publisher}{Novosibirsk: Izd-vo SO RAN}.
\newblock \bibinfo{note}{(In Russ.)}.
\bibitem[{Lannes(2017)}]{Lannes_2017}
\bibinfo{author}{Lannes, D.}, \bibinfo{year}{2017}.
\newblock \bibinfo{title}{On the dynamics of floating structures}.
\newblock \bibinfo{journal}{Annals of PDE} \bibinfo{volume}{3},
  \bibinfo{pages}{Paper~11}.
\newblock \DOIprefix\doi{https://doi.org/10.1007/s40818-017-0029-5}.
\bibitem[{Lee and Chwang(2000)}]{Lee_Chwang_2000}
\bibinfo{author}{Lee, M.M.}, \bibinfo{author}{Chwang, A.T.},
  \bibinfo{year}{2000}.
\newblock \bibinfo{title}{Scattering and radiation of water waves by permeable
  barriers}.
\newblock \bibinfo{journal}{Phys. Fluids} \bibinfo{volume}{12 (1)},
  \bibinfo{pages}{54--65}.
\newblock \DOIprefix\doi{https://doi.org/10.1063/1.870284}.
\bibitem[{Lin(2006)}]{Lin_2006}
\bibinfo{author}{Lin, P.}, \bibinfo{year}{2006}.
\newblock \bibinfo{title}{A multiple-layer $\sigma$-coordinate model for
  simulation of wave–structure interaction}.
\newblock \bibinfo{journal}{Comput. Fluids} \bibinfo{volume}{35 (2)},
  \bibinfo{pages}{147--167}.
\newblock \DOIprefix\doi{10.1016/j.compfluid.2004.11.008}.
\bibitem[{Losada et~al.(1992)Losada, Losada and Roldan}]{Losada}
\bibinfo{author}{Losada, I.J.}, \bibinfo{author}{Losada, M.A.},
  \bibinfo{author}{Roldan, A.J.}, \bibinfo{year}{1992}.
\newblock \bibinfo{title}{Propagation of oblique incident waves past rigid
  vertical thin barriers}.
\newblock \bibinfo{journal}{Appl. Ocean Res.} \bibinfo{volume}{14 (3)},
  \bibinfo{pages}{191--199}.
\newblock \DOIprefix\doi{https://doi.org/10.1016/0141-1187(92)90014-B}.
\bibitem[{Lu and Wang(2015)}]{Lu_Wang_2015}
\bibinfo{author}{Lu, X.}, \bibinfo{author}{Wang, K.H.}, \bibinfo{year}{2015}.
\newblock \bibinfo{title}{Modeling a solitary wave interaction with a fixed
  floating body using an integrated analytical–numerical approach}.
\newblock \bibinfo{journal}{Ocean Eng.} \bibinfo{volume}{109},
  \bibinfo{pages}{691--704}.
\newblock \DOIprefix\doi{https://doi.org/10.1016/j.oceaneng.2015.09.050}.
\bibitem[{Madsen et~al.(2002)Madsen, Bingham and Liu}]{Madsen_Bingham_Liu_2002}
\bibinfo{author}{Madsen, P.A.}, \bibinfo{author}{Bingham, H.B.},
  \bibinfo{author}{Liu, H.}, \bibinfo{year}{2002}.
\newblock \bibinfo{title}{A new {B}oussinesq method for fully nonlinear waves
  from shallow to deep water}.
\newblock \bibinfo{journal}{J. Fluid Mech.} \bibinfo{volume}{462},
  \bibinfo{pages}{1--30}.
\newblock \DOIprefix\doi{https://doi.org/10.1017/S0022112002008467}.
\bibitem[{Mavrakos(1985)}]{Mavrakos_1985}
\bibinfo{author}{Mavrakos, S.A.}, \bibinfo{year}{1985}.
\newblock \bibinfo{title}{Wave loads on a stationary floating bottomless
  cylindrical body with finite wall thickness}.
\newblock \bibinfo{journal}{Appl. Ocean Res.} \bibinfo{volume}{7 (4)},
  \bibinfo{pages}{213--224}.
\newblock \DOIprefix\doi{https://doi.org/10.1016/0141-1187(85)90028-8}.
\bibitem[{Mei and Black(1969)}]{Mei_Black_1969}
\bibinfo{author}{Mei, C.C.}, \bibinfo{author}{Black, J.L.},
  \bibinfo{year}{1969}.
\newblock \bibinfo{title}{Scattering of surface waves by rectangular obstacles
  in waters of finite depth}.
\newblock \bibinfo{journal}{J. Fluid Mech.} \bibinfo{volume}{38},
  \bibinfo{pages}{499--511}.
\newblock \DOIprefix\doi{https://doi.org/10.1017/S0022112069000309}.
\bibitem[{Nudner et~al.(2017)Nudner, Semenov, Khakimzyanov and
  Shokina}]{Nudner_2017}
\bibinfo{author}{Nudner, I.S.}, \bibinfo{author}{Semenov, K.K.},
  \bibinfo{author}{Khakimzyanov, G.S.}, \bibinfo{author}{Shokina, N.Y.},
  \bibinfo{year}{2017}.
\newblock \bibinfo{title}{Investigations of the long marine waves interaction
  with the structures protected by the vertical barriers}.
\newblock \bibinfo{journal}{Fundamental and Applied Hydrophysics}
  \bibinfo{volume}{10 (4)}, \bibinfo{pages}{31--43}.
\newblock \DOIprefix\doi{https://doi.org/10.7868/S2073667317040037}.
  \bibinfo{note}{(In Russ.)}.
\bibitem[{Nudner et~al.(2019)Nudner, Semenov, Lebedev, Khakimzyanov and
  Zakharov}]{Nudner_2019}
\bibinfo{author}{Nudner, I.S.}, \bibinfo{author}{Semenov, K.K.},
  \bibinfo{author}{Lebedev, V.V.}, \bibinfo{author}{Khakimzyanov, G.S.},
  \bibinfo{author}{Zakharov, Y.N.}, \bibinfo{year}{2019}.
\newblock \bibinfo{title}{Numerical model of the hydrowave laboratory for
  studying the interaction of sea waves with hydrotechnical structures}.
\newblock \bibinfo{journal}{J. Computat. Technol.} \bibinfo{volume}{24 (1)},
  \bibinfo{pages}{86--105}.
\newblock \DOIprefix\doi{10.25743/ICT.2019.24.1.007}. \bibinfo{note}{(In
  Russ.)}.
\bibitem[{Orzech et~al.(2016)Orzech, Shi, Veeramony, Bateman, Calantoni and
  Kirby}]{Orzech_2016}
\bibinfo{author}{Orzech, M.D.}, \bibinfo{author}{Shi, F.},
  \bibinfo{author}{Veeramony, J.}, \bibinfo{author}{Bateman, S.},
  \bibinfo{author}{Calantoni, J.}, \bibinfo{author}{Kirby, J.T.},
  \bibinfo{year}{2016}.
\newblock \bibinfo{title}{Incorporating floating surface objects into a fully
  dispersive surface wave model}.
\newblock \bibinfo{journal}{Ocean Model.} \bibinfo{volume}{102},
  \bibinfo{pages}{14--26}.
\newblock \DOIprefix\doi{https://doi.org/10.1016/j.ocemod.2016.04.007}.
\bibitem[{Palagina and Kha\-kimzyanov(2019)}]{Palagina_2019}
\bibinfo{author}{Palagina, A.A.}, \bibinfo{author}{Kha\-kimzyanov, G.S.},
  \bibinfo{year}{2019}.
\newblock \bibinfo{title}{Numerical simulation of surface waves in a basin with
  moving impermeable boundaries}.
\newblock \bibinfo{journal}{J. Computat. Technol.} \bibinfo{volume}{24 (4)},
  \bibinfo{pages}{70--107}.
\newblock \DOIprefix\doi{10.25743/ICT.2019.24.4.006}. \bibinfo{note}{(In
  Russ.)}.
\bibitem[{Park et~al.(2001)Park, Kim and Miyata}]{Park_etal_2001}
\bibinfo{author}{Park, J.C.}, \bibinfo{author}{Kim, M.H.},
  \bibinfo{author}{Miyata, H.}, \bibinfo{year}{2001}.
\newblock \bibinfo{title}{Three-dimensional numerical wave tank simulations on
  fully nonlinear wave–current–body interactions}.
\newblock \bibinfo{journal}{J. Mar. Sci. Technol.} \bibinfo{volume}{6},
  \bibinfo{pages}{70–82}.
\newblock \DOIprefix\doi{https://doi.org/10.1007/s773-001-8377-2}.
\bibitem[{Pelinovsky(1996)}]{Pelinovsky_1996}
\bibinfo{author}{Pelinovsky, E.N.}, \bibinfo{year}{1996}.
\newblock \bibinfo{title}{Tsunami Wave Hydrodynamics}.
\newblock \bibinfo{publisher}{Nizhny Novgorod: Institute Applied Physics
  Press}.
\newblock \bibinfo{note}{(In Russ.)}.
\bibitem[{Shokin et~al.(2019)Shokin, Gusiakov, Kikhtenko and
  Chubarov}]{Shokin_2019}
\bibinfo{author}{Shokin, Y.I.}, \bibinfo{author}{Gusiakov, V.K.},
  \bibinfo{author}{Kikhtenko, V.A.}, \bibinfo{author}{Chubarov, L.B.},
  \bibinfo{year}{2019}.
\newblock \bibinfo{title}{A methodology for mapping tsunami hazards and its
  implementation for the {Far Eastern coast of the Russian Federation}}.
\newblock \bibinfo{journal}{Dokl. Earth Sc.} \bibinfo{volume}{489},
  \bibinfo{pages}{1444--1448}.
\newblock \DOIprefix\doi{10.1134/S1028334X19120092}.
\bibitem[{Sun et~al.(2015)Sun, Wang, Wu and Khoo}]{Sun_etal_2015}
\bibinfo{author}{Sun, J.L.}, \bibinfo{author}{Wang, C.Z.}, \bibinfo{author}{Wu,
  G.X.}, \bibinfo{author}{Khoo, B.C.}, \bibinfo{year}{2015}.
\newblock \bibinfo{title}{Fully nonlinear simulations of interactions between
  solitary waves and structures based on the finite element method}.
\newblock \bibinfo{journal}{Ocean Eng.} \bibinfo{volume}{108},
  \bibinfo{pages}{202–215}.
\newblock \DOIprefix\doi{http://dx.doi.org/10.1016/j.oceaneng.2015.08.007}.
\bibitem[{Zheleznyak(1985)}]{Zheleznyak_1985}
\bibinfo{author}{Zheleznyak, M.I.}, \bibinfo{year}{1985}.
\newblock \bibinfo{title}{Influence of long waves on vertical obstacles}, in:
  \bibinfo{editor}{Pelinovsky, E.N.} (Ed.), \bibinfo{booktitle}{Tsunami
  climbing a beach: {C}ollection of scientific papers}.
  \bibinfo{publisher}{Gorky: Institute Applied Physics Press}, p.
  \bibinfo{pages}{122–140}.
\newblock \bibinfo{note}{(In Russ.)}.

\end{thebibliography}


\end{document}